\documentclass[aps,twocolumn,nofootinbib,groupedaddress,superscriptaddress,longbibliography,notitlepage]{revtex4-2}

\usepackage{amsmath,amssymb}
\usepackage[normalem]{ulem}
\usepackage{graphicx}
\usepackage{longtable,multirow}
\usepackage{mathtools}
\usepackage{dsfont}
\usepackage{amsfonts}
\usepackage[dvipsnames]{xcolor}
\usepackage{url}
\usepackage{bbold}
\usepackage[colorlinks=true,linkcolor=WildStrawberry,citecolor=WildStrawberry,urlcolor=RoyalBlue,hypertexnames=false]{hyperref}
\newcommand{\nocontentsline}[3]{}
\newcommand{\tocless}[2]{\bgroup\let\addcontentsline=\nocontentsline#1{#2}\egroup}

\def\ba#1\ea{\begin{align}#1\end{align}}
\def\bg#1\eg{\begin{gather}#1\end{gather}}
\def\bpm{\begin{pmatrix}}
\def\epm{\end{pmatrix}}
\def\bbm{\begin{bmatrix}}
\def\ebm{\end{bmatrix}}

\newcommand{\nn}{\nonumber \\ }
\newcommand{\bb}[1]{{\mathbf #1}}

\newcommand{\bx}{\bb x}
\newcommand{\bk}{\bb k}
\newcommand{\bp}{\bb p}

\newcommand{\bR}{\bb R}
\newcommand{\bD}{\bb \Delta}

\newcommand{\cm}{\overline}
\newcommand{\mc}[1]{\mathcal{#1}}

\newcommand{\dg}{\dagger}

\newcommand{\sg}{\sigma}

\newcommand{\vep}{\varepsilon}
\newcommand{\ep}{\epsilon}
\newcommand{\Z}{\mathbb{Z}}
\newcommand{\R}{\mathbb{R}}

\newcommand{\C}{\mathbb{C}}

\newcommand{\ket}[1]{|#1\rangle}
\newcommand{\bra}[1]{\langle#1|}
\newcommand{\brk}[2]{\langle#1|#2\rangle}

\newcommand{\vev}[1]{\langle#1\rangle}

\newcommand{\gsk}{{\ket{\rm GS}}}
\newcommand{\gsb}{{\bra{\rm GS}}}

\newcommand{\themecol}[1]{{\color{WildStrawberry} #1}}

\newcommand{\ourtitle}{Exciton Alchemy: Chern Excitons from Trivial Bands}

\allowdisplaybreaks

\begin{document}
\title{\ourtitle}

\author{Yoonseok Hwang}
\author{Henry Davenport}
\author{Frank Schindler}
\affiliation{Blackett Laboratory, Imperial College London, London SW7 2AZ, United Kingdom}

\begin{abstract}
Exciton topology is commonly inherited from the topology of the underlying electronic bands.
Recent theoretical work, however, has shown that the exciton Chern number can in general receive an additional contribution from the topology of the exciton envelope wave function, allowing, in principle, interaction-induced topological excitons even when the constituent electronic bands are topologically trivial.
Here, we provide an explicit realization of this case by constructing a two-dimensional exciton model with topologically trivial conduction and valence bands that nevertheless hosts a Chern exciton diagnosed by inversion symmetry.
Starting from a real-space limit of exponentially localized Wannier states for the conduction and valence bands, we identify the essential ingredients responsible for the emergent exciton topology and formulate a simple construction recipe.
Our work demonstrates that interactions alone can generate nontrivial exciton topology, independent of the topology of the underlying electronic bands, and establishes a general framework for designing interaction-induced topological excitons.
\end{abstract}

\maketitle

\let\oldaddcontentsline\addcontentsline
\renewcommand{\addcontentsline}[3]{}

\themecol{\it Introduction---}
Topology provides a unified framework for understanding robust properties of quantum states, ranging from quantized bulk responses to protected boundary states in crystalline systems~\cite{hasan2010colloquium,qi2011topological,chiu2016classification,shiozaki2016topology,shiozaki2017topological,kruthoff2017topological,wang2016hourglass,bradlyn2016beyond,wieder2018wallpaper,hwang2023magnetic}.
In electronic band theory, symmetry eigenvalues at high-symmetry momenta provide powerful constraints on band topology and form the basis of symmetry-based approaches such as topological quantum chemistry and symmetry indicators~\cite{fu2007topological,hughes2011inversion,turner2012quantized,fang2012bulk,alexandradinata2014wilson,bradlyn2017topological,cano2018building,elcoro2021magnetic,hwang2026srsi,hwang2026building,po2017symmetry,watanabe2018structure,song2018quantitative,khalaf2018symmetry}.
More broadly, understanding the interplay between topology and symmetry beyond single-particle band theory has become an important theme across interacting quantum systems.

Excitons are bound states of electrons and holes and provide a natural platform to explore the interplay of topology and interactions~\cite{mott1961transition}.
Various forms of exciton topology have recently been proposed and investigated, including Chern excitons, shift excitons, and related topological excitonic phases~\cite{yao2008berry,haber2023maximally,wu2017topological,chen2017chiral,gong2017chiral,blason2020exciton,kwan2021exciton,xie2024theory,xie2024long,froese2025topological,kwan2025textured,cai2026continuum,eto2026odd,paiva2024shift,lozano2025optical,qiu2025quantum,davenport2026composite,davenport2024interaction,davenport2026exciton,davenport2026berry,thompson2025topologically,jankowski2025excitonic,hwang2026stable}.
More recently, increasing attention has been devoted to understanding how crystalline symmetries constrain and characterize exciton topology~\cite{davenport2024interaction,davenport2026exciton,davenport2026berry,thompson2025topologically,jankowski2025excitonic,hwang2026stable,nalabothula2026symmetries}.
These developments suggest that symmetry can provide a systematic framework for understanding and engineering topological exciton states, analogous to its role in electronic band theory.

Among these developments, Chern excitons have attracted particular interest. 
Existing studies have found Chern excitons exclusively in electronic bands  with nonzero Chern numbers~\cite{wu2017topological,chen2017chiral,gong2017chiral,blason2020exciton,kwan2021exciton,xie2024theory,xie2024long,froese2025topological,kwan2025textured,cai2026continuum,eto2026odd,paiva2024shift,lozano2025optical}.
By contrast, recent work has identified a new class of excitons in one dimension (1D), called shift excitons, whose nontrivial Berry phase arises from interactions alone, even though the underlying bands are topologically trivial~\cite{davenport2024interaction,davenport2026exciton}.
This example illustrates more generally that exciton topology need not be determined solely by the topology of the constituent electronic bands.
In principle, additional topology carried by the exciton envelope wave function can also contribute~\cite{paiva2024shift,hwang2026stable}.

In Ref.~\cite{hwang2026stable}, we developed a symmetry-based classification of exciton topology and the corresponding stable-zero patterns at high-symmetry momenta (HSMs) in the presence of inversion or rotation symmetry.
Within this symmetry-based classification, Chern excitons in trivial electronic bands are not forbidden by symmetry constraints~\cite{hwang2026stable}.
This case, where the Chern numbers of the conduction and valence bands vanish, would provide the clearest realization of genuinely interaction-induced exciton topology.
However, no explicit microscopic construction realizing this possibility has been reported.
It is therefore unclear whether such excitons can exist at all, or whether they can be generated systematically within trivial electronic structures.

\begin{figure}
\centering
\includegraphics[width=0.47\textwidth]{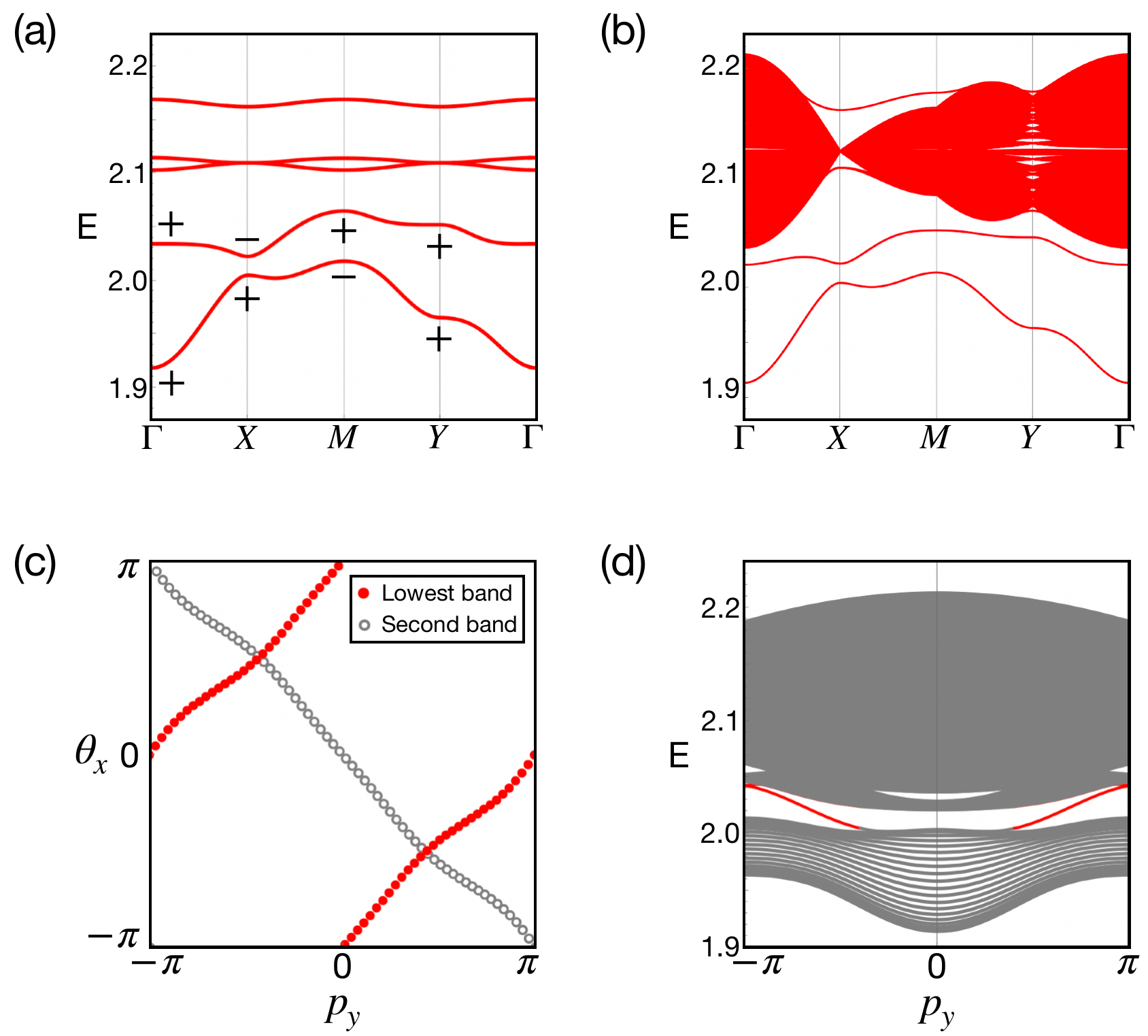}
\caption{(a) Exciton band structure of the minimal $5 \times 5$ model after introducing the additional terms $t_{x+y}=0.02$ and $U'_x=0.015$.
A full bulk gap is opened above the lowest exciton band.
(b) Exciton band structure with the enlarged and more realistic $441 \times 441$ basis.
Additional exciton continua emerge at higher energies, while the low-energy bands remain essentially unchanged.
(c) Wilson-loop spectra of the lowest and second exciton bands in (b), corresponding to Chern numbers $C_{\rm exc}=+1$ and $-1$, respectively.
(d) Ribbon spectrum with 20 layers along the open $x$ direction.
The chiral exciton edge states traversing the bulk gap are highlighted in red.}
\label{fig:detail}
\end{figure}

In this work, we demonstrate the possibility of interaction-induced Chern excitons by constructing an explicit microscopic model with topologically trivial underlying electronic bands.
For concreteness, we consider inversion-symmetric systems and start from the atomic-insulator limit of the electronic bands, where both the conduction and valence bands are topologically trivial.
This construction combines the real-space framework of Ref.~\cite{davenport2024interaction} for 1D shift excitons with the symmetry-based classification developed in Ref.~\cite{hwang2026stable}, extending the former to interaction-induced Chern excitons.
Beyond providing an explicit existence proof, we identify the essential ingredients responsible for the emergent topology and formulate a construction recipe, paving the way for the systematic realization of interaction-induced topological excitons.
Figure~\ref{fig:detail} summarizes the resulting model and its key topological properties.

\begin{figure*}[t]
\centering
\includegraphics[width=0.98\textwidth]{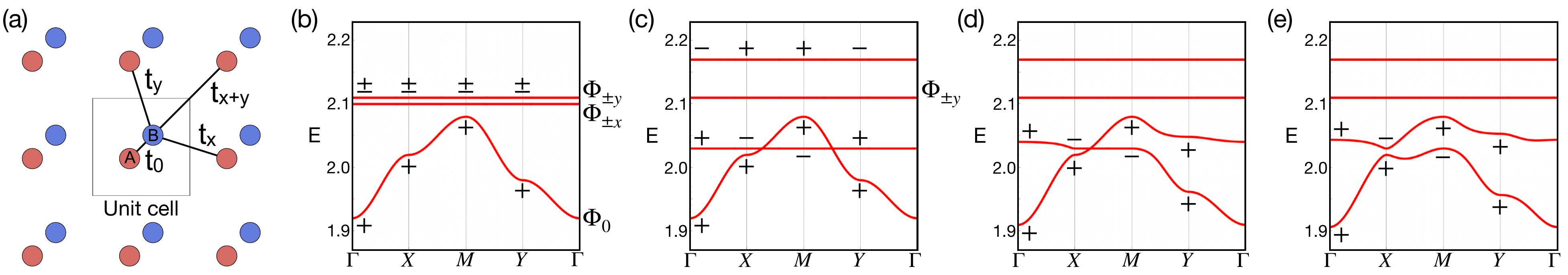}
\caption{
(a) Unit cell and hopping geometry of the model with $A$ and $B$ sublattices.
We consider $t_0=1$, $t_x=0.025$, $t_{x+y}=0.02$, and $t_y=0$.
These hopping amplitudes can be naturally realized on an oblique lattice as the first-, second-, third-, and fifth-nearest-neighbor hoppings, respectively~\cite{supple}.
For notational convenience, however, we choose the primitive lattice vectors to be $(1,0)$ and $(0,1)$ throughout this work.
(b) Exciton band structure in the atomic limit ($t_0=1$) after introducing the density-density interactions $(U_0,U_x,U_y)=(0.125,0.10,0.06)$.
The inversion eigenstates $\Phi_0$, $\Phi_{\pm x}$, and $\Phi_{\pm y}$, together with their inversion eigenvalues at the four HSMs, are indicated.
At every HSM, the $\Phi_{\pm x}$ and $\Phi_{\pm y}$ doublets each consist of one even- and one odd-parity state.
(c) Interaction $V_x=0.07$, inducing the desired band inversion while leaving an accidental nodal ring.
The $\Phi_{\pm y}$ doublet and the highest band remain exactly flat throughout panels (c)–(e).
(d) Additional hopping $t_x=0.025$, leaving only nodal points on the $X$-$M$ line.
(e) Additional interaction $V'=0.03$, yielding an isolated lowest exciton band.
The inversion eigenvalues indicate that the two lowest bands are Chern bands.}
\label{fig:steps}
\end{figure*}

\themecol{\it Excitons in the relative distance basis---}
Our construction builds on the real-space description of excitons introduced in Ref.~\cite{davenport2024interaction}, which we briefly summarize here and generalize to 2D.
We consider the simplest real-space setting in which an interaction-induced Chern exciton can arise.
Specifically, we consider a 2D system with inversion or two-fold rotation symmetry $\mc I$ and two sublattices, $A$ and $B$, in each unit cell [Fig.~\ref{fig:steps}(a)].
For concreteness, we adopt the minimal two-sublattice setting, although our construction itself is not restricted to two-sublattice models~\cite{supple}.
$\mc I$ exchanges the two sublattices and is therefore represented by $\sg_x$ in the $(A,B)$ basis.
We take the conduction and valence bands to be topologically trivial, so that exponentially localized Wannier states can be chosen~\cite{brouder2007exponential,marzari2012maximally}.
We start from the atomic limit with Hamiltonian,
\ba
H_0 = - t_0 \sum_\bR \, \left( c^\dg_{\bR, A} \, c_{\bR, B} + c^\dg_{\bR, B} \, c_{\bR, A} \right),
\label{eq:H0}
\ea
where $\bR$ labels the unit cells.
We set the overall energy scale $t_0$ to unity.
The valence and conduction Wannier states are the even and odd combinations of the two sublattices, with creation operators, $c^\dagger_{\bb R,n} = \frac{1}{\sqrt2} (c^\dg_{\bR, A} \pm c^\dg_{\bR, B})$, where the plus and minus signs correspond to the valence $n = v$ and conduction $n = c$ bands respectively.
They transform as $s$- and $p$-like orbitals under $\mc I$, respectively.

We define the real-space exciton basis, $\ket{\bD,\bR} = c^\dg_{\bR,c} c_{\bR-\bD,v} \gsk$, where $\gsk$ is the filled valence-band ground state within the conduction-valence Hilbert space.
Here, $\bD$ denotes the electron-hole separation and $\bR$ the overall position of the exciton.
Fourier transforming with respect to $\bR$ gives $\ket{\bD,\bp}
= \frac{1}{\sqrt N} \sum_\bR \, e^{i \bp \cdot \bR} \ket{\bD, \bR}$,
so that $\bp$ is the total exciton momentum.
Further details are given in the Supplemental Material~\cite{supple}.

The exciton spectrum is obtained from the projected Hamiltonian, $\mc H_{\bD, \bD'} (\bR) = \bra{\bD, \bR} H \ket{\bD', \bb 0} - \delta_{\bD, \bD'} \delta_{\bR, \bb 0} \gsb H \gsk$, where translation symmetry has been used~\cite{davenport2024interaction,supple}.
Its momentum-space representation is $\mc H_{\bD, \bD'} (\bp) = \sum_\bR \, e^{-i \bp \cdot \bR} \, \mc H_{\bD, \bD'} (\bR)$.
Inversion symmetry plays a central role in the construction.
In the exciton basis, it reverses both the relative displacement and the exciton momentum, $\mc I \ket{\bD, \bp} = \ket{-\bD, -\bp}$, up to an overall phase that can be absorbed without loss of generality~\cite{supple}.

\themecol{\it Inversion-guided exciton band inversion---}
We now construct an interaction-induced Chern exciton by introducing interactions on top of the atomic limit $H_0$.
Throughout the construction, inversion symmetry serves as the guiding principle, since inversion eigenvalues at HSMs diagnose the Chern number modulo two~\cite{fu2007topological,hughes2011inversion,turner2012quantized,fang2012bulk,alexandradinata2014wilson}.
Since the lowest exciton states are dominated by small electron-hole separations~\cite{supple}, we retain only the relative displacements $\bD \in \left\{\bb 0, \hat x, -\hat x, \hat y, -\hat y \right\}$, where $\hat a$ denotes the unit vector along the $a=x,y$ direction.
This yields a five-band projected exciton Hamiltonian at each $\bp$.
With this 5-dimensional basis, inversion takes the block-diagonal form,
\bg
\mc I = \mathbb{1}_{1 \times 1} \oplus \sg_x \oplus \sg_x,
\label{eq:inversion_action}
\eg
where $\sg_x$ is the first $2 \times 2$ Pauli matrix.

The projected Hamiltonian from $H_0$ is simply $\mc H_{\bD, \bD'} (\bp) = 2 \delta_{\bD, \bD'}$.
We first introduce the repulsive density-density interactions $U_0 \, n_{\bR, A} \, n_{\bR, B}$ and $U_a \, n_{\bR + \hat a, A} \, n_{\bR, B}$ ($a=x,y$), where $n_{\bR, A/B}$ is the electron number operator, summation over $\bR$ is understood, and $U_{0, a}>0$.
The resulting projected Hamiltonian remains diagonal, with $\mc H (\bp) = {\rm Diag} \left( \vep_0 (\bp), \vep_x, \vep_x, \vep_y, \vep_y \right)$ where $\vep_0 (\bp) = 2 - \frac{U_x}{2} \cos p_x -\frac{U_y}{2} \cos p_y$ and $\vep_a = 2+U_0-\frac{U_a}{4}$ for $a=x,y$~\cite{supple}.
It is convenient to introduce the inversion eigenbasis, $\Phi_0 = (1,0,0,0,0)^T$, $\Phi_{\pm x} = \frac{1}{\sqrt2} (0,1,\pm 1,0,0)^T$, and $\Phi_{\pm y} = \frac{1}{\sqrt2} (0,0,0,1,\pm 1)^T$, which are simultaneous eigenstates of the projected Hamiltonian with energies $\vep_0 (\bp)$, $\vep_x$, and $\vep_y$, respectively.
As shown in Fig.~\ref{fig:steps}(b), for $(U_0, U_x, U_y) = (0.125, 0.10, 0.06)$, the spectrum consists of a dispersive $\Phi_0$ band and two flat doublets $\Phi_{\pm x}$ and $\Phi_{\pm y}$.

Using Eq.~\eqref{eq:inversion_action}, one immediately finds that $\Phi_0$, $\Phi_{+x}$, and $\Phi_{+y}$ have inversion eigenvalue $+1$, whereas $\Phi_{-x}$ and $\Phi_{-y}$ have inversion eigenvalue $-1$ at the HSMs $\Gamma=(0,0)$, $X=(\pi,0)$, $Y=(0,\pi)$, and $M=(\pi,\pi)$, as illustrated in Fig.~\ref{fig:steps}(b).
According to the inversion-eigenvalue formula for the Chern number in the absence of time-reversal symmetry (TRS)~\cite{fu2007topological,hughes2011inversion,turner2012quantized,fang2012bulk,alexandradinata2014wilson} and its excitonic generalization~\cite{hwang2026stable}, an exciton band has Chern number $C_{\rm exc}=1 \bmod 2$ if its total number of odd inversion eigenvalues over all HSMs is odd.
The correspondence with the stable-zero classification of Ref.~\cite{hwang2026stable} is discussed in the SM~\cite{supple}.
Our task therefore reduces to inducing a band inversion between $\Phi_0$ and one of the odd-parity states.

The parameters $U_x$ and $U_y$ determine the relative energy of the two doublets with respect to $\Phi_0$.
For our chosen parameters, the exciton bands are ordered as $\Phi_0$, $\Phi_{\pm x}$, and $\Phi_{\pm y}$.
Since $\Phi_0$ reaches its maximum energy at the $M$ point, the desired band inversion is achieved by selectively lowering the odd-parity state $\Phi_{-x}$ within the $\Phi_{\pm x}$ doublet.
To this end, we introduce the symmetry-allowed interaction, $V_x \, c^\dg_{\bR + \hat x, A} \, c_{\bR, A} \, c^\dg_{\bR + \hat x, B} \, c_{\bR, B} + h.c.$, where $V_x \in \R$ and $h.c.$ denotes the Hermitian conjugate.
This term simultaneously translates the $A$ and $B$ electrons by one lattice spacing along $\hat x$.
Projecting onto the exciton basis, this interaction contributes
\bg
0_{1 \times 1} \oplus \bpm 0 & - e^{-i p_x} V_x \\ - e^{i p_x} V_x & 0 \epm \oplus 0_{2 \times 2},
\label{eq:five_H}
\eg
where $0_{n \times n}$ denotes the $n \times n$ zero matrix.
Thus, the $V_x$ term acts exclusively within the $\Phi_{\pm x}$ subspace, hybridizing $\Phi_{+x}$ and $\Phi_{-x}$ while leaving the remaining states unaffected.
As shown in Fig.~\ref{fig:steps}(c), setting $V_x=0.07$ drives the lower branch below $\Phi_0$ at the $M$ point, producing the desired band inversion.
The spectrum in Fig.~\ref{fig:steps}(c) nevertheless remains gapless because of a nodal ring around $M$.

\themecol{\it Emergent time-reversal symmetry---}
The origin of the remaining degeneracy in the exciton spectrum is an emergent spinless TRS of the projected exciton Hamiltonian.
As shown in the SM~\cite{supple}, the projected Hamiltonian obtained from arbitrary inversion-preserving hopping terms $t c^\dg_{\bR, \alpha} \, c_{\bR', \beta}$ and density-density interactions $U n_{\bR, \alpha} \, n_{\bR', \beta}$ ($\alpha,\beta=A,B$) always exhibits this spinless TRS.
Consequently, an inversion eigenvalue pattern that would imply $C_{\rm exc}=1 \bmod 2$ in the absence of TRS instead gives rise to symmetry-protected nodal points~\cite{kim2015dirac,ono2018unified,song2018diagnosis}.

Only the nodal points between the two lowest bands are enforced by the inversion eigenvalues.
The nodal ring in Fig.~\ref{fig:steps}(c) is therefore accidental.
We next introduce the leading symmetry-allowed hopping processes beyond the atomic limit, $t_x \, c^\dg_{\bR + \hat x, A} \, c_{\bR, B} + h.c.$ ($t_x \in \R$).
As shown in Fig.~\ref{fig:steps}(d), where we set $t_x=0.025$, the previously flat second band acquires dispersion, and the accidental nodal ring is reduced to the nodal points on the $X$-$M$ line.
Along this line, the two lowest bands are formed by $\Phi_0$ and $\Phi_{-x}$.
Their hybridization is generated by the $t_x$ contribution to the projected Hamiltonian [Eq.~\eqref{eq:t_break} below with $V'=0$] at generic momenta, but vanishes identically at $p_x=\pi$.

To remove the remaining nodal points, it is necessary to break the TRS.
To this end, we introduce the inversion-symmetric term, $i V' \, c^\dg_{\bR - \hat x, A} \, c_{\bR + \hat y, A}$ $c^\dg_{\bR + \hat y, A} \, c_{\bR, B} + i V' \, c^\dg_{\bR + \hat x, B} \, c_{\bR - \hat y, B} \, c^\dg_{\bR - \hat y, B} \, c_{\bR, A} + h.c.$, where $V' \in \R$.
The necessity of this interaction becomes evident after projection onto the exciton basis.
Together with the $t_x$ hopping terms, it contributes to
\bg
\bpm 0_{1 \times 1} & h_\bp \\
h_\bp^\dg & 0_{4 \times 4} \epm,
\label{eq:t_break}
\eg
where $h_\bp = \left( f_\bp, f_{-\bp}, 0, 0 \right)$ with $f_\bp = -\frac{t_x}{2} (1 + e^{i p_x}) + i \frac{V'}{4} (e^{i p_x + i p_y} + e^{-i p_y})$.
The $V'$ term generates the missing hybridization between $\Phi_0$ and $\Phi_{-x}$ on the $X$-$M$ line and gaps the remaining nodal points.
As shown in Fig.~\ref{fig:steps}(e), where we set $V'=0.03$, a full gap opens between the two lowest exciton bands.

The desired isolated lowest exciton band has thus been successfully constructed.
Its inversion eigenvalues at $(\Gamma,X,Y,M)$ are $(+,+,+,-)$, implying $C_{\rm exc}=1 \bmod 2$.
Interestingly, the second band is also identified as a Chern exciton band from its inversion eigenvalues $(+,-,+,+)$.
In this work, however, we focus exclusively on the topology of the lowest band.

\themecol{\it Gapped Chern exciton and topological signatures---}
The minimal construction above identifies the essential ingredients required to generate a Chern exciton band.
To obtain a fully gapped realization supporting well-defined chiral edge states, we further include the symmetry-allowed terms,
\bg
t_{x+y} \, ( c^\dg_{\bR + \hat x + \hat y, A} \, c_{\bR, B} + c^\dg_{\bR, B} \, c_{\bR + \hat x + \hat y, A} )
\nn
+ U'_x \, (n_{\bR + \hat x, A} \, n_{\bR, A} + n_{\bR + \hat x, B} \, n_{\bR, B}),
\eg
where $t_{x+y}, U'_x \in \R$, whose projected forms are given in the SM~\cite{supple}.
These terms are introduced solely to enlarge the bulk gap and do not modify the band topology established above.

The dominant effect of the $t_{x+y}$ term is to lower the lowest exciton band near the $M$ point while leaving its energy near the $X$ point almost unchanged.
We therefore tune its strength until the band extrema at $M$ and $X$ become nearly degenerate, at which point the minimum gap between the two lowest bands is controlled by the $X$ point.
The interaction $U'_x$ then increases this remaining gap without a gap closing, thereby opening a full bulk gap above the lowest exciton band.
The parameters used throughout this section are $t_{x+y}=0.02$ and $U'_x=0.015$.
The resulting bulk spectrum is shown in Fig.~\ref{fig:detail}(a).

To verify that the construction is not an artifact of the truncated basis, we next enlarge the exciton basis to include all relative displacements $\bD = (\Delta_x, \Delta_y)$ satisfying $|\Delta_{x,y}| \le 10$, corresponding to a $441 \times 441$ projected Hamiltonian at each $\bp$.
The full bulk gap above the lowest exciton band is preserved, while the enlarged basis gives rise to a much richer exciton continuum [Fig.~\ref{fig:detail}(b)].
Note that the remaining degeneracies near the $M$ point merely reflect our minimal choice of hopping terms.

We next compute the Wilson-loop spectra of the isolated exciton bands.
Let $\phi_{n,\bD}(\bp)$ denote the $\bD$th component of the normalized eigenvector of $\mc H_{\bD,\bD'}(\bp)$ corresponding to the $n$th exciton band, normalized as $\sum_\bD \, |\phi_{n,\bD} (\bp)|^2 = 1$.
For each fixed $p_y$, we calculate the Wilson loop
\bg
\mc W^{(n)} (p_y) =
\prod_{p_x=0}^{2\pi-\delta p} \, \sum_\bD \,  \phi^*_{n,\bD} (p_x', p_y) \phi_{n,\bD} (p_x,p_y)
\label{eq:wl}
\eg
where $p_x'=p_x+\delta p$ with the momentum spacing $\delta p$ along the discretized $p_x$ direction.
The Wilson-loop spectrum consists of $\theta^{(n)}_x (p_y) = \arg \mc W^{(n)} (p_y) \in [-\pi,\pi)$ plotted as a function of $p_y \in [-\pi,\pi)$.
See the SM for details~\cite{supple}.
Figure~\ref{fig:detail}(c) displays the Wilson-loop spectra of the first and second exciton bands, respectively.
They exhibit the expected nontrivial winding corresponding to $C_{\rm exc}=+1$ and $-1$, respectively~\cite{alexandradinata2014wilson,davenport2026exciton}, in agreement with the inversion-eigenvalue analysis presented above.
Importantly, Eq.~\eqref{eq:wl} involves only the exciton envelope wave function.
The resulting Chern topology is therefore generated by interactions, independently of the topology of the underlying electronic bands.

Finally, Fig.~\ref{fig:detail}(d) shows the spectrum of a ribbon geometry with $L_x=20$ layers and open boundaries along the $x$ direction.
We retain the same range of relative coordinates $\bD$ as in the bulk calculation, but restrict the exciton basis $\ket{\bD,\bR}=c^\dg_{\bR,c} \, c_{\bR-\bD,v} \gsk$ to configurations satisfying $1 \le R_x \le L_x$, and $1 \le R_x-\Delta_x \le L_x$, so that both the electron and hole remain inside the ribbon.
The resulting spectrum exhibits chiral edge states traversing the bulk gap, confirming the nontrivial topology of the lowest exciton band.

\themecol{\it Discussion---}
In this work, we presented an explicit microscopic construction of interaction-induced Chern excitons from topologically trivial electronic bands.
Starting from the atomic limit, we identified the essential ingredients required to engineer an isolated Chern exciton band, including band inversion, emergent symmetries, and correlated interactions.
Our construction therefore provides both an existence proof and a symmetry-guided framework for designing topological excitons.

An intriguing aspect of our construction is the emergent spinless TRS of the projected exciton Hamiltonian.
Understanding when such emergent symmetries arise, and whether other lattice geometries or space groups admit Chern excitons through more conventional hopping processes and density-density interactions, remains an interesting direction for future work.
It is likewise important to identify microscopic mechanisms that generate the correlated interactions introduced here in realistic systems.

Finally, since excitons are electrically neutral quasiparticles, the physical consequences of exciton topology are expected to differ fundamentally from those of electronic Chern insulators.
Understanding these consequences remains an important direction for future work.
\\

\themecol{\it Note added---}
A very recent paper, Ref.~\cite{zheng2026topologicalchargetransferexcitons} also finds Chern excitons in trivial electronic bands. Here, the nontrivial topology arises from a real space embedding instead of a specific interaction.

\begin{acknowledgments}
This work was supported by a UKRI Future Leaders Fellowship MR/Y017331/1. HD acknowledges support from the Engineering and Physical Sciences Research Council (grant number EP/W524323/1).
\end{acknowledgments}

\bibliography{refs.bib}

\let\addcontentsline\oldaddcontentsline
\clearpage

\onecolumngrid
\begin{center}
\textbf{\large Supplemental Material: \\
\ourtitle}
\end{center}

\setcounter{section}{0}
\setcounter{figure}{0}
\setcounter{equation}{0}
\renewcommand{\thefigure}{S\arabic{figure}}
\renewcommand{\theequation}{S\arabic{equation}}
\renewcommand{\thesection}{S\arabic{section}}

\tableofcontents
\hfill \\

\section{Excitons in the relative distance basis}
\label{app:real_space}

\subsection{Wannier and exciton basis}
\label{subapp:real_basis}
We begin by introducing the real-space description of excitons, following Ref.~\cite{davenport2024interaction}.
Since the underlying conduction ($c$) and valence ($v$) bands are assumed to be topologically trivial, exponentially localized Wannier states can be chosen.
Let $c^\dg_\bk$ and $v^\dg_\bk$ denote the creation operators for the conduction and valence bands at crystal momentum $\bk$, respectively.
Their corresponding Wannier creation operators are defined by
\bg
c^\dg_\bR = \frac{1}{\sqrt N} \sum_\bk \, e^{- i \bk \cdot \bR} c^\dg_\bk,
\quad
v^\dg_\bR = \frac{1}{\sqrt N} \sum_\bk \, e^{- i \bk \cdot \bR} v^\dg_\bk,
\quad
c^\dg_\bk = \frac{1}{\sqrt N} \sum_\bR \, e^{i \bk \cdot \bR} c^\dg_\bR,
\quad
v^\dg_\bk = \frac{1}{\sqrt N} \sum_\bR \, e^{i \bk \cdot \bR} v^\dg_\bR.
\label{seq:wannier}
\eg
where $N$ is the number of unit cells, $\bR$ labels the unit cells.
For notational simplicity, we write $c^\dg_\bR$ ($v^\dg_\bR$) instead of $c^\dg_{\bR,c}$ ($c^\dg_{\bR,v}$) used in the main text.
We denote the intracell positions of the conduction- and valence-band Wannier states  as $\bx_c$ and $\bx_v$ so that the Wannier states are localized at $\bR+\bx_c$ and $\bR+\bx_v$.

We then define the real-space exciton basis as
\bg
\ket{\bD, \bR} = c^\dg_\bR \, v_{\bR - \bD} \gsk,
\label{seq:basis_real}
\eg
where $\gsk = \prod_\bR \, v^\dg_\bR \ket{0}$ is the filled-valence reference state in the active conduction-valence Hilbert space, with the valence band completely filled and the conduction band empty.
Here $\ket{0}$ denotes the vacuum.
The electron-hole separation in this basis is $\bD + \bx_c - \bx_v$.
Since the intracell displacement $\bx_c - \bx_v$ is fixed, the relative electron-hole separation is uniquely characterized by the lattice vector $\bD$.
We next perform a Fourier transformation and introduce the basis labeled by the relative displacement $\bD$ and the total crystal momentum $\bp$ of the exciton:
\bg
\ket{\bD, \bp} = \frac{1}{\sqrt N} \sum_\bR \, e^{i \bp \cdot \bR} \ket{\bD, \bR},
\quad
\ket{\bD, \bR} = \frac{1}{\sqrt N} \sum_\bp \, e^{-i \bp \cdot \bR} \ket{\bD, \bp}.
\label{seq:basis_partial1}
\eg
The interpretation of $\bp$ as the exciton total momentum is established in Sec.~\ref{app:momentum}.

We now define the projected Hamiltonian.
Let $\hat H$ denote the full Hamiltonian of the system.
The projected Hamiltonian in the real-space exciton basis is defined as
\bg
\mc H_{\bD, \bD'} (\bR - \bR')
= \bra{\bD, \bR} \hat H \ket{\bD', \bR'}
- \delta_{\bD, \bD'} \delta_{\bR, \bR'} \gsb H \gsk,
\label{seq:projH_r}
\eg
where the second term subtracts the reference-state energy.

The dependence on $\bR-\bR'$ follows from translation symmetry.
Let $\hat t_{\bR_0}$ denote the translation operator by a lattice vector $\bR_0$.
Since $\hat t_{\bR_0} \ket{\bD, \bR} = \ket{\bD, \bR + \bR_0}$ and $\hat t_{\bR_0} \, \hat H \, \hat t_{\bR_0}^{-1} = \hat H$, the right-hand side of Eq.~\eqref{seq:projH_r} can be rewritten as
\bg
\bra{\bD, \bR - \bR_0} \hat H \ket{\bD', \bR' - \bR_0}
- \delta_{\bD, \bD'} \delta_{\bR-\bR', \bb 0} \gsb H \gsk.
\eg
Choosing $\bR_0=\bR'$, we conclude that $\mc H_{\bD, \bD'}$ depends only on $\bR-\bR'$.

Next, we define the projected Hamiltonian in the basis $\ket{\bD, \bp}$:
\bg
\mc H_{\bD, \bD'} (\bp) = \bra{\bD, \bp} \hat H \ket{\bD', \bp} - \delta_{\bD, \bD'} \gsb \hat H \gsk.
\eg
From Eq.~\eqref{seq:basis_partial1}, one immediately obtains
\ba
\mc H_{\bD, \bD'} (\bp)
= \sum_{\bR} \, e^{-i \bp \cdot \bR} \, \mc H_{\bD, \bD'} (\bR).
\label{seq:projH_h}
\ea
In this basis, the eigenvalue problem is
\bg
\sum_{\bD'} \, \mc{H}_{\bD, \bD'} (\bp) \, \phi_{n,\bD'} (\bp)
= E_n (\bp) \phi_{n,\bD} (\bp),
\eg
where $n$ labels the exciton bands.
The eigenvectors are normalized as
\bg
\sum_\bD \, \left( \phi_{n,\bD} (\bp) \right)^* \phi_{n',\bD} (\bp) = \delta_{n,n'}.
\eg
The coefficients $\phi_{n,\bD} (\bp)$ are the exciton envelope wave functions in the $\ket{\bD, \bp}$ basis, such that the corresponding exciton eigenstate $\ket{\Phi_n (\bp)}$ is
\bg
\ket{\Phi_n (\bp)} =
\sum_\bD \, \phi_{n,\bD} (\bp) \, \ket{\bD, \bp}.
\label{seq:exc_state}
\eg
It is then convenient to introduce the projected Hamiltonian operator
\bg
\hat{\mc H} (\bp) = \sum_{\bD, \bD'} \, \ket{\bD, \bp} \mc H_{\bD, \bD'} (\bp) \bra{\bD', \bp},
\label{seq:projHop}
\eg
which satisfies $\mc H (\bp) \ket {\Phi_n (\bp)} = E_n (\bp) \ket{\Phi_n (\bp)}$.

\subsection{Inversion symmetry}
\label{subapp:inversion}
We derive the representation of inversion symmetry $\mc I$ in the various bases.
We begin by defining its action on the Wannier basis:
\bg
\hat {\mc I} \, c^\dg_\bR \, \hat {\mc I}^{-1} = \lambda_c (\bb 0) \, c^\dg_{-\bR - 2 \bx_c},
\quad
\hat {\mc I} \, v^\dg_\bR \, \hat {\mc I}^{-1} = \lambda_v (\bb 0) \, v^\dg_{-\bR - 2 \bx_v}.
\label{seq:inv_wannier}
\eg
The coefficients $\lambda_l (\bb 0)$ ($l=c,v$) are the inversion eigenvalues of the corresponding Wannier states.
More generally, for a nondegenerate Wannier state in an inversion-symmetric system, $\lambda_l (\bb 0)=+1$ ($-1$) corresponds to even- (odd-) parity, i.e. an $s$-like ($p$-like) Wannier state.
The meaning of the argument $\bb 0$ will become clear after introducing the momentum-space representation below.

The intracell positions $\bx_l$ satisfy $2 \bx_l \in \Lambda_{\rm Latt}$, where $\Lambda_{\rm Latt}$ denotes the Bravais lattice.
Indeed, for a nondegenerate band, inversion maps the Wannier center at $\bx_l$ to $-\bx_l$, which must represent the same Wannier state modulo a lattice vector.
The transformation law in Eq.~\eqref{seq:inv_wannier} can then be understood geometrically.
For example, $c^\dg_\bR$ creates the conduction-band Wannier state centered at $\bR + \bx_c$.
Under inversion, $\bR + \bx_c$ is mapped to $-\bR - \bx_c = (-\bR - 2 \bx_c) + \bx_c$.
Since $2 \bx_c \in \Lambda_{\rm Latt}$, $-\bR - 2 \bx_c$ is again a valid unit-cell label.
Hence the transformed creation operator is $c^\dg_{-\bR-2\bx_c}$ up to the inversion eigenvalue $\lambda_c (\bb 0)$.
The transformation of $v^\dg_\bR$ is understood analogously.

The momentum-space transformation law for $c^\dg_\bk$ (and similarly for $v^\dg_\bk$) is
\ba
\hat {\mc I} c^\dg_\bk \hat {\mc I}^{-1}
= \hat {\mc I} \left[ \frac{1}{\sqrt N} \sum_\bR \, e^{i \bk \cdot \bR} c^\dg_\bR \right] \hat {\mc I}^{-1}
= \frac{1}{\sqrt N} \sum_\bR \, \lambda_c(\bb 0) e^{i \bk \cdot \bR} c^\dg_{-\bR - 2 \bx_c}
= \lambda_c (\bb 0) e^{- 2i \bk \cdot \bx_c} c^\dg_{-\bk},
\ea
where we used Eqs.~\eqref{seq:wannier} and \eqref{seq:inv_wannier}.
We therefore define the inversion sewing matrix
\bg
\lambda_l (\bk) = \lambda_l (\bb 0) e^{- 2i \bk \cdot \bx_l},
\label{seq:inv_sewing}
\eg
for $l=c,v$.
Note that $\lambda_l (-\bk) \lambda_l (\bk) = 1$,which follows directly from $\hat{\mc I}^2=1$.
At inversion-invariant momenta, the sewing matrix $\lambda_l (\bk)$ reduce to the inversion eigenvalues $\pm 1$.
Combining Eqs.~\eqref{seq:inv_wannier} and \eqref{seq:inv_sewing}, we obtain
\bg
\hat {\mc I} \, c^\dg_\bk \, \hat {\mc I}^{-1} = \lambda_c (\bk) \, c^\dg_{-\bk},
\quad
\hat {\mc I} \, v^\dg_\bk \, \hat {\mc I}^{-1} = \lambda_v (\bk) \, v^\dg_{-\bk}.
\label{seq:inv_band}
\eg

We next derive the representation of inversion symmetry in the exciton bases $\ket{\bD, \bR}$ and $\ket{\bD, \bp}$.
Define $\bD_{\mc I} = - \bD + 2 \bx_v - 2 \bx_c$ and let $\lambda_{\rm GS} \in \{+1,-1\}$ denote the inversion eigenvalue of the reference ground state, $\hat {\mc I} \gsk = \lambda_{\rm GS} \gsk$.
Then,
\bg
\hat {\mc I} \ket{\bD, \bR}
= (\hat {\mc I} \, c^\dg_\bR \, \hat {\mc I}^{-1}) \, (\hat {\mc I} \, v_{\bR-\bD} \, \hat {\mc I}^{-1}) \, \hat {\mc I} \gsk
= \ket{ \bD_I, -\bR - 2 \bx_c} \, \lambda_{\rm GS} \lambda_c (\bb 0) \lambda_v (\bb 0)^{-1}.
\label{seq:inv_op1}
\eg
A Fourier transformation gives
\bg
\hat {\mc I} \ket{\bD, \bp}
= \ket{\bD_I, -\bp} \, e^{- 2i \bp \cdot \bx_c} \lambda_{\rm GS} \lambda_c (\bb 0) \lambda_v (\bb 0)^{-1}.
\label{seq:inv_op2}
\eg
This immediately yields the representation of inversion symmetry in the $\ket{\bD, \bp}$ basis:
\bg
\hat U_{\mc I} (\bp)
= \sum_\bD \, \ket{ \bD_I, -\bp} \, e^{-2 i \bp \cdot \bx_c} \lambda_{\rm GS} \lambda_c (\bb 0) \lambda_v (\bb 0)^{-1} \, \bra{\bD, \bp}.
\label{seq:inv_op3}
\eg
Motivated by this representation, we define the exciton inversion sewing matrix by
\bg
\lambda_{\rm GS}^{-1} \, \hat U_{\mc I} (\bp) \ket{\Phi (\bp)} = \ket{\Phi (-\bp)} \, \lambda_{\rm exc} (\bp).
\label{seq:exc_inversion1}
\eg
The exciton-band index $n$ is omitted for simplicity.
At inversion-invariant momenta, $\lambda_{\rm exc} (\bp)$ become inversion eigenvalues.
The prefactor $\lambda_{\rm GS}^{-1}$ in Eq.~\eqref{seq:exc_inversion1} removes the contribution from $\gsk$, so that $\lambda_{\rm exc} (\bp)$ represents the inversion eigenvalue relative to the reference ground state.
The above relation can be written explicitly in the $\ket{\bD,\bp}$ basis as
\bg
e^{-2 i \bp \cdot \bx_c} \, \lambda_c (\bb 0) \lambda_v (\bb 0)^{-1} \, \phi_{\bD} (\bp) = \lambda_{\rm exc} (\bp) \, \phi_{\bD_{\mc I}} (-\bp).
\label{seq:exc_inversion2}
\eg

Finally, inversion symmetry constrains the projected Hamiltonian as
\bg
\hat U_{\mc I} (\bp) \hat {\mc H} (\bp) \hat U_{\mc I} (\bp)^{-1} = \hat {\mc H} (-\bp),
\eg
or equivalently,
\bg
\mc H_{\bD, \bD'} (\bp) = \mc H_{\bD_{\mc I}, \bD'_{\mc I}} (-\bp).
\eg

\section{Application to the present model}
\label{app:model}

\subsection{Projected Hamiltonian}
\label{subapp:model_H}
We now specialize the general formalism to the model considered in the main text.
The lattice geometry is shown in Fig.~\ref{sfig:lattice}(a), where the primitive lattice vectors, sublattice positions, and hopping processes are illustrated.
The atomic-limit Hamiltonian is
\bg
\hat H_0 = - t_0 \sum_\bR \, \left( c^\dg_{\bR, A} \, c_{\bR, B} + h.c. \right),
\eg
with $t_0=1$.
Introducing the Wannier basis
\bg
v^\dg_\bR = \frac{1}{\sqrt 2} (c^\dg_{\bR,A} + c^\dg_{\bR,B}),
\quad
c^\dg_\bR = \frac{1}{\sqrt 2} (c^\dg_{\bR,A} - c^\dg_{\bR,B}),
\label{seq:model_basis}
\eg
the Hamiltonian becomes
\bg
-t_0 \sum_\bR \, \left( n_{\bR,v} - n_{\bR,c} \right),
\eg
where $n_{\bR,v} = v^\dg_\bR \, v_\bR$ and $n_{\bR,c} = c^\dg_\bR \, c_\bR$ are the valence- and conduction-band number operators, respectively.
This shows that the chosen Wannier basis diagonalizes the atomic-limit Hamiltonian.
The valence and conduction Wannier orbitals therefore correspond to $s$-like and $p$-like orbitals localized at the unit-cell center, respectively.
Conversely, one may start from a chosen Wannier basis and construct a parent Hamiltonian realizing it.
This viewpoint naturally extends to systems with multiple sublattices per unit cell, as discussed in Sec.~\ref{app:multiband}.

First, we determine the representation of inversion symmetry $\mc I$ for the present model.
Inversion exchanges the $A$ and $B$ sublattices, so that $\mc I$ is represented by the Pauli matrix $\sg_x$ in the sublattice basis.
Accordingly, $\hat {\mc I} \, v^\dg_\bR \, \hat{\mc I}^{-1} = + v^\dg_{-\bR}$ and $\hat {\mc I} \, c^\dg_\bR \, \hat{\mc I}^{-1} = - c^\dg_{-\bR}$.
Comparing with Eq.~\eqref{seq:inv_wannier}, we identify $\lambda_v (\bb 0) = +1$, $\lambda_c (\bb 0) = -1$, and $\bx_c = \bx_v = \bb 0$.
Using Eqs.~\eqref{seq:basis_partial1} and \eqref{seq:inv_op3}, the inversion operator in the $\ket{\bD, \bp}$ basis is therefore
\bg
\hat U_{\mc I} (\bp)
= - \lambda_{\rm GS} \sum_\bD \, \ket{-\bD, -\bp} \bra{\bD, \bp}.
\eg
In the main text, the overall sign factor $-\lambda_{\rm GS}$ is absorbed into the definition of the inversion operator.
Consequently, inversion simply exchanges $(\bD, \bp) \leftrightarrow (-\bD, -\bp)$.

We now write the full microscopic Hamiltonian introduced in the main text:
\ba
\hat H =& - t_0 \sum_\bR \, \left( c^\dg_{\bR, A} \, c_{\bR, B} + h.c. \right)
+ t_x \sum_\bR \, \left( c^\dg_{\bR + \bb a_1, A} \, c_{\bR, B} + h.c. \right)
+ t_{x+y} \sum_ \bR \, \left( c^\dg_{\bR + \bb a_1 + \bb a_2, A} \, c_{\bR, B} + h.c. \right)
\nn
&+ U_0 \sum_\bR \, n_{\bR, A} \, n_{\bR, B}
+ U_x \sum_\bR \, n_{\bR + \bb a_1, A} \, n_{\bR, B}
+ U_y \sum_\bR \, n_{\bR + \bb a_2, A} \, n_{\bR, B}
\nn
&+ U'_x \sum_\bR \, \left( n_{\bR + \bb a_1, A} \, n_{\bR, A} + n_{\bR + \bb a_1, B} \, n_{\bR, B} \right)
+ V_x \sum_\bR \, \left( c^\dg_{\bR + \bb a_1, A} \, c_{\bR, A} \, c^\dg_{\bR + \bb a_1, B} \, c_{\bR, B} + h.c. \right)
\nn
&+ V' \sum_\bR \, \left( i c^\dg_{\bR - \bb a_1, A} \, c_{\bR + \bb a_2, A} \, c^\dg_{\bR + \bb a_2, A} \, c_{\bR, B} + i c^\dg_{\bR + \bb a_1, B} \, c_{\bR - \bb a_2, B} \, c^\dg_{\bR - \bb a_2, B} \, c_{\bR, A} + h.c. \right).
\label{seq:fullmodel}
\ea
The parameters used throughout the main text are $t_0=1$, $t_x=0.025$, $t_{x+y}=0.02$, $t_y=0$, $U_0=0.125$, $U_x=0.10$, $U_y=0.06$, $U'_x=0.015$, $V_x=0.07$, and $V'=0.03$.

Now, we evaluate the projected Hamiltonian $\mc H_{\bD, \bD'} (\bp)$ from the microscopic Hamiltonian.
According to Eq.~\eqref{seq:projH_h}, we need to evaluate $\bra{\bD, \bR} \hat H \ket{\bD', \bb 0}$ and $\gsb \hat H \gsk$ where $\gsk = \prod_\bR \, v^\dg_\bR \ket{0}$.
Let $\hat h$ denote an arbitrary term in $\hat H$.
Its contribution to $\mc H_{\bD,\bD'}(\bp)$ is then given by
\bg
h_{\bD, \bD'} (\bp)
= \left[ \sum_{\bR} \, e^{-i \bp \cdot \bR} \, \bra{\bD, \bR} \hat H \ket{\bD', \bb 0} \right] - \delta_{\bD, \bD'} \gsb \hat H \gsk.
\label{seq:projH_p}
\eg
where the Fourier transformation follows from Eqs.~\eqref{seq:projH_r} and \eqref{seq:projH_h}.

\tocless{\subsubsection{The $t_0$ term}}{}
The $t_0$ term is
\bg
\hat h = -t_0 \sum_\bR \, \left( v^\dg_\bR \, v_\bR - c^\dg_\bR \, c_\bR \right).
\eg
Since the reference ground state has the valence band completely filled, $v^\dg_\bR \gsk = 0$ and $c_\bR \gsk = 0$.
Therefore,
\bg
\gsb v^\dg_\bR \, v_\bR \gsk
= \gsb (1-v_\bR \, v^\dg_\bR) \gsk = 1,
\quad
\gsb c^\dg_\bR \, c_\bR \gsk = 0.
\eg
It follows immediately that $\gsb \hat h \gsk = -t_0 \sum_\bR \, 1 = - t_0 N$.
The matrix element $\bra{\bD, \bR} \hat h \ket{\bD', \bb 0}$ is obtained from the definition of the $\ket{\bD, \bR}$ basis [Eq.~\eqref{seq:basis_real}] as
\bg
\bra{\bD, \bR} \hat h \ket{\bD', \bb 0}
= - t_0 \sum_{\bR'} \, \gsb v^\dg_{\bR - \bD} \, c_\bR \, v^\dg_{\bR'} \, v_{\bR'} \, c^\dg_{\bb 0} \, v_{-\bD'} \gsk
+ t_0 \sum_{\bR'} \, \gsb v^\dg_{\bR - \bD} \, c_\bR \, c^\dg_{\bR'} \, c_{\bR'} \, c^\dg_{\bb 0} \, v_{-\bD'} \gsk.
\label{seq:term_example1}
\eg
Consider first the first term in Eq.~\eqref{seq:term_example1}.
Using the anticommutation relation,
\bg
- t_0 \sum_{\bR'} \, \gsb v^\dg_{\bR - \bD} \, v^\dg_{\bR'} \, v_{\bR'} \, v_{-\bD'} \, c_\bR \, c^\dg_{\bb 0} \gsk
= - t_0 \delta_{\bR, \bb0} \sum_{\bR'} \, \gsb v^\dg_{-\bD} \, v^\dg_{\bR'} \, v_{\bR'} \, v_{-\bD'} \gsk,
\eg
where we used $c_\bR \gsk = 0$.
Moving the remaining valence creation operators $v^\dg$ to the right and using $v^\dg_\bR \gsk = 0$, we obtain
\bg
v^\dg_{-\bD} \, v^\dg_{\bR'} \, v_{\bR'} \, v_{-\bD'} \gsk = \left[ \delta_{-\bD, -\bD'} \delta_{\bR', \bR'} - \delta_{-\bD, \bR'} \delta_{\bR', -\bD'} \right] \gsk.
\label{seq:term_example2}
\eg
Therefore, the first term in Eq.~\eqref{seq:term_example1} evaluates to $-t_0 (N-1) \delta_{\bD, \bD'} \delta_{\bR, \bb 0}$.

In Eq.~\eqref{seq:term_example2}, the Kronecker deltas $\delta_{-\bD, -\bD'} \delta_{\bR', \bR'}$ and $\delta_{-\bD, \bR'} \delta_{\bR', -\bD'}$ may be rewritten as
\bg
\gsb v^\dg_{-\bD} \, v_{-\bD'} \gsk \gsb v^\dg_{\bR'} \, v_{\bR'} \gsk
\quad \text{and} \quad
\gsb v^\dg_{-\bD} \, v_{\bR'} \gsk \gsb v^\dg_{\bR'} \, v_{-\bD'} \gsk,
\eg
respectively.
This motivates the use of the standard Wick theorem.
We define $\vev{\mc O} := \gsb \mc O \gsk$.
For the $\gsk = \prod_\bR \, v^\dg_\bR \ket{0}$, the nonvanishing contractions are
\bg
\vev{v^\dg_\bR \, v_{\bR'}}
= \delta_{\bR, \bR'},
\quad
\vev{c_\bR \, c^\dg_{\bR'}} = \delta_{\bR, \bR'}.
\label{seq:wick}
\eg

Applying Wick's theorem to the second term in Eq.~\eqref{seq:term_example1}, we obtain
\bg
t_0 \sum_{\bR'} \, \gsb v^\dg_{\bR - \bD} \, c_\bR \, c^\dg_{\bR'} \, c_{\bR'} \, c^\dg_{\bb 0} \, v_{-\bD'} \gsk.
= t_0 \sum_{\bR'} \, \vev{v^\dg_{\bR-\bD} \, v_{-\bD'}} \left[ \vev{c_\bR \, c^\dg_{\bR'}} \vev{c_{\bR'} \, c^\dg_{\bb 0}} + \vev{c_\bR \, c^\dg_{\bb 0}} \vev{c^\dg_{\bR'} \, c_{\bR'}} \right].
\eg
Using Eq.~\eqref{seq:wick}, this reduces to
\bg
t_0 \sum_{\bR'} \, \delta_{\bR-\bD,-\bD'} \delta_{\bR,\bR'} \delta_{\bR', \bb 0}
= t_0 \sum_{\bR'} \, \delta_{\bD,\bD'} \delta_{\bR, \bb 0} \delta_{\bR', \bb 0}
= t_0 \delta_{\bD,\bD'} \delta_{\bR, \bb 0}.
\eg
Therefore,
\bg
\bra{\bD, \bR} \hat h \ket{\bD', \bb 0} = t_0 (2-N) \delta_{\bR, \bb 0} \delta_{\bD,\bD'},
\quad
\vev{\hat h} = - t_0 N.
\eg
Substituting this result into Eq.~\eqref{seq:projH_p}, we finally obtain
\bg
h_{\bD,\bD'} (\bp) = 2 t_0 \delta_{\bD,\bD'}.
\eg
\\

\tocless{\subsubsection{The $t_x$ and $t_{x+y}$ terms}}{}
The $t_x$ and $t_{x+y}$ terms have the same structure.
Suppressing the hopping amplitude, we write
\bg
\hat h = \sum_\bR \, \left( c^\dg_{\bR + \bb X, A} \, c_{\bR, B} + h.c. \right)
\eg
where $\bb X = \bb a_1$ for the $t_x$ term and $\bb a_1 + \bb a_2$ for the $t_{x+y}$ term.
Expressed in the Wannier basis,
\bg
\hat h = \frac{1}{2} \sum_\bR \, \left( v^\dg_{\bR+\bb X} \, v_\bR - c^\dg_{\bR+\bb X} \, c_\bR \right)
+ \frac{1}{2} \sum_\bR \, \left( c^\dg_{\bR+\bb X} \, v_\bR - v^\dg_{\bR+\bb X} \, c_\bR \right) + h.c.
\eg
The second term changes the number of conduction and valence electrons and therefore does not contribute to either $\bra{\bD, \bR} \hat h \ket{\bD', \bb 0}$ or $\gsb \hat h \gsk$.
Hence, only the first term and its Hermitian conjugate need to be retained:
\bg
\frac{1}{2} \sum_\bR \, v^\dg_{\bR+\bb X} \, v_\bR
- \frac{1}{2} \sum_\bR \, c^\dg_{\bR+\bb X} \, c_\bR + h.c.
\label{seq:term_example3}
\eg

Let us evaluate the contribution of this term to the projected Hamiltonian $h_{\bD, \bD'} (\bp)$.
We first compute $\vev{\hat h}$.
The first and second terms in Eq.~\eqref{seq:term_example3} give
\bg
\frac{1}{2} \vev{v^\dg_{\bR+\bb X} \, v_\bR}
= \frac{1}{2} \delta_{\bb X, \bb 0},
\quad
- \frac{1}{2} \vev{c^\dg_{\bR+\bb X} \, c_\bR} = 0.
\eg
Therefore, $\vev{\hat h} = N \delta_{\bb X, \bb 0}$.
Next, we evaluate $\bra{\bD, \bR} \hat h \ket{\bD', \bb 0}$, which contains two contributions:
\bg
\frac{1}{2} \sum_{\bR'} \, \vev{v^\dg_{\bR - \bD} \, c_\bR \, v^\dg_{\bR'+\bb X} \, v_{\bR'} \, c^\dg_{\bb 0} \, v_{-\bD'}}
- \frac{1}{2} \sum_{\bR'} \, \vev{v^\dg_{\bR - \bD} \, c_\bR \, c^\dg_{\bR'+\bb X} \, c_{\bR'} \, c^\dg_{\bb 0} \, v_{-\bD'}}
\eg
The first contribution is
\ba
& \frac{1}{2} \sum_{\bR'} \, \vev{c_\bR \, c^\dg_{\bb 0}} \left( \vev{v^\dg_{\bR-\bD} \, v_{-\bD'}} \vev{v^\dg_{\bR'+\bb X} \, v_{\bR'}} - \vev{v^\dg_{\bR-\bD} \, v_{\bR'}} \vev{v^\dg_{\bR'+\bb X} \, v_{-\bD'}} \right)
\nn
=& \frac{1}{2} \sum_{\bR'} \, \delta_{\bR, \bb 0} (\delta_{\bD,\bD'} \delta_{\bb X, \bb 0} - \delta_{\bD, -\bR'} \delta_{\bD, \bD'+\bb X})
= \frac{1}{2} \delta_{\bR, \bb 0} (N \delta_{\bD,\bD'} \delta_{\bb X, \bb 0} - \delta_{\bD,\bD'+\bb X}).
\ea
The second contribution is obtained similarly:
\bg
- \frac{1}{2} \sum_{\bR'} \, \vev{v^\dg_{\bR - \bD} \, c_\bR \, c^\dg_{\bR'+\bb X} \, c_{\bR'} \, c^\dg_{\bb 0} \, v_{-\bD'}}
=- \frac{1}{2} \delta_{\bR, \bb X} \delta_{\bD,\bD'+\bb X}.
\eg

The contribution from the Hermitian conjugate terms in Eq.~\eqref{seq:term_example3} is obtained by $(\bD,\bD',\bR) \leftrightarrow (\bD',\bD,-\bR)$ followed by complex conjugation.
Indeed, if $\hat h = \hat g + \hat g^\dg$ with $\hat g$ translationally invariant,
\bg
\bra{\bD, \bR} \hat g^\dg \ket{\bD', \bb 0} = (\bra{\bD', \bb 0} \hat g \ket{\bD, \bR})^* = (\bra{\bD', \bb -\bR} \hat g \ket{\bD, \bb 0})^*.
\eg
Combining all contributions, we obtain
\bg
\bra{\bD, \bR} \hat h \ket{\bD', \bb 0}
= N \delta_{\bR, \bb 0} \delta_{\bD,\bD'} \delta_{\bb X, \bb 0}
- \frac{1}{2} (\delta_{\bR, \bb X} + \delta_{\bR, \bb 0}) \delta_{\bD,\bD'+\bb X}
- \frac{1}{2} (\delta_{\bR, -\bb X} + \delta_{\bR, \bb 0}) \delta_{\bD+\bb X,\bD'},
\quad
\vev{\hat h} = N \delta_{\bb X, \bb 0}.
\eg
Substituting this result into Eq.~\eqref{seq:projH_p}, we obtain
\bg
h_{\bD,\bD'} (\bp) = - \frac{1}{2} (e^{-i \bp \cdot \bb X} + 1) \delta_{\bD, \bD'+\bb X} - \frac{1}{2} (e^{i \bp \cdot \bb X} + 1) \delta_{\bD+\bb X, \bD'}.
\eg

\tocless{\subsubsection{Other terms}}{}
The remaining terms can be evaluated in the same manner using Wick's theorem.
We simply list the results below.
\\

\paragraph*{a. The $U_0$, $U_x$, and $U_y$ terms:}
The first three density-density interactions can be written in the unified form
\bg
\hat h = \sum_\bR \, n_{\bR + \bb X, A} \, n_{\bR, B},
\eg
where $\bb X = \bb 0$, $\bb a_1$, $\bb a_2$ correspond to the $U_0$, $U_x$, and $U_y$ terms, respectively.
Their contribution is
\bg
h_{\bD,\bD'} (\bp)
= \delta_{\bb X, \bb 0} \delta_{\bD, \bD'}
- \frac{1}{2} \cos (\bp \cdot \bb X) \, \delta_{\bD, \bb 0} \delta_{\bD', \bb 0}
- \frac{1}{4} \delta_{\bD, \bb X} \delta_{\bD', \bb X}
- \frac{1}{4} \delta_{\bD, -\bb X} \delta_{\bD', -\bb X}.
\eg

\paragraph*{b. The $U'_x$ term:}
The $U_x'$ interaction can be written as
\bg
\hat h = \sum_\bR \, \left( n_{\bR + \bb X, A} \, n_{\bR, A} + n_{\bR + \bb X, B} \, n_{\bR, B} \right),
\eg
where $\bb X = \bb a_1$.
Its contribution is
\bg
h_{\bD,\bD'} (\bp)
= \cos (\bp \cdot \bb X) \, \delta_{\bD, \bb 0} \delta_{\bD', \bb 0}
- \frac{1}{2} \delta_{\bD, \bb X} \delta_{\bD', \bb X}
- \frac{1}{2} \delta_{\bD, -\bb X} \delta_{\bD', -\bb X}.
\eg

\paragraph*{c. Other interaction terms:}
The $V_x$ interaction can be written as
\bg
\hat h = \sum_\bR \left( c^\dg_{\bR + \bb X, A} \, c_{\bR, A} \, c^\dg_{\bR + \bb X, B} \, c_{\bR, B} + c^\dg_{\bR, A} \, c_{\bR + \bb X, A} \, c^\dg_{\bR, B} \, c_{\bR + \bb X, B} \right),
\eg
where $\bb X = \bb a_1$.
Its contribution is
\bg
h_{\bD,\bD'} (\bp)
= 2 \delta_{\bb X, \bb0} \delta_{\bD, \bD'} - e^{-i \bp \cdot \bb X} \delta_{\bD, \bb X} \delta_{\bD', -\bb X} - e^{i \bp \cdot \bb X} \delta_{\bD, -\bb X} \delta_{\bD', \bb X}.
\eg
Finally, we consider
\bg
\hat h = i V' \sum_\bR \, \left(c^\dg_{\bR - \bb a_1, A} \, c_{\bR + \bb a_2, A} c^\dg_{\bR + \bb a_2, A} \, c_{\bR, B}
+ c^\dg_{\bR + \bb a_1, B} \, c_{\bR - \bb a_2, B} \, c^\dg_{\bR - \bb a_2, B} \, c_{\bR, A} \right) + h.c..
\eg
Although inversion symmetry and Hermiticity allow $V' \in \C$, we restrict to $V' \in \R$ for simplicity in the present model.
The explicit factor of $i$ nevertheless breaks the emergent spinless time-reversal symmetry explained in Sec.~\ref{subapp:trs}.
Its contribution is
\bg
h_{\bD,\bD'} (\bp)
= \frac{iV'}{4} \big[
(e^{-i p_y} + e^{i p_x + i p_y}) \delta_{\bD, \bb 0} \delta_{\bD', \hat x}
- (e^{i p_y} + e^{-i p_x - i p_y}) \delta_{\bD, \hat x} \delta_{\bD', \bb 0}
\nn
+ (1+ e^{i p_x}) \delta_{\bD, \hat y} \delta_{\bD', \hat x + \hat y}
- (1 + e^{-i p_x}) \delta_{\bD, \hat x + \hat y} \delta_{\bD', \hat y} \big]
+ \frac{i V'}{4} \Big[ (\bD, \bD', \bp) \leftrightarrow (-\bD, -\bD', -\bp) \Big].
\eg
The notation $[(\bD, \bD',\bp) \leftrightarrow (-\bD, -\bD', -\bp)]$ denotes the preceding expression but with the replacement $(\bD, \bD', \bp) \rightarrow (-\bD, -\bD', -\bp)$.

\tocless{\subsubsection{Continuum excitations beyond the truncated exciton basis}}{}
In the main text, we constructed our model using a truncated exciton basis consisting of the five relative coordinates, $\bD \in \{\bb 0, +\hat x, -\hat x, +\hat y, -\hat y\}$.
Here we explain why this truncation is justified when the terms introduced beyond the atomic limit are sufficiently small.
To this end, we consider the simplified model containing only the hopping $t_0$ and the interaction terms $U_0$, $U_x$, and $U_y$, which form the starting point of our construction in the main text.
In this case, the spectrum can be obtained exactly.
Comparing the truncated and infinite-$\bD$ bases shows that extending the basis primarily introduces additional high-energy continuum states, while the low-energy bound states are already captured by the truncated basis.

In this case, the projected Hamiltonian is given by
\bg
\mc H_{\bD, \bD'} (\bp)
= 2 t_0 \delta_{\bD, \bD'} + U_0 \left( \delta_{\bD, \bD'} - \delta_{\bD, \bb 0} \delta_{\bD', \bb 0} \right) - \sum_{a=x,y} \, \frac{U_a}{4} \left( 2 \cos p_a \, \delta_{\bD, \bb 0} \delta_{\bD', \bb 0}
+ \delta_{\bD, \hat a} \delta_{\bD', \hat a}
+ \delta_{\bD, -\hat a} \delta_{\bD', -\hat a} \right).
\label{seq:H_5x5}
\eg
We first consider the truncated basis.
The Hamiltonian is diagonal,
\bg
\mc H (\bp) = {\rm Diag} \left( 2t_0 - \frac{U_x}{2} \cos p_x - \frac{U_y}{2} \cos p_y, 2t_0 + U_0 - \frac{U_x}{4}, 2t_0 + U_0 - \frac{U_x}{4}, 2t_0 + U_0 - \frac{U_y}{4}, 2t_0 + U_0 - \frac{U_y}{4} \right),
\eg
written in the basis $\ket{\bD, \bp}$, where the relative coordinates $\bD$ are ordered as $\{\bb 0, +\hat x, -\hat x, +\hat y, -\hat y\}$.
The corresponding eigenstates $\phi (\bp)$ and eigenvalues $E (\bp)$ are
\ba
& \phi_1 (\bp) = (1,0,0,0,0) \quad \text{and} \quad E_1 (\bp) = 2 t_0 - \frac{U_x}{2} \cos p_x - \frac{U_y}{2} \cos p_y,
\nn
& \phi_2 (\bp) = \frac{1}{\sqrt 2}(0,1,1,0,0) \quad \text{and} \quad E_2 (\bp) = 2 t_0 +U_0 - \frac{U_x}{4},
\nn
& \phi_3 (\bp) = \frac{1}{\sqrt 2}(0,1,-1,0,0) \quad \text{and} \quad E_3 (\bp) = 2 t_0 +U_0 - \frac{U_x}{4},
\nn
& \phi_4 (\bp) = \frac{1}{\sqrt 2}(0,0,0,1,1) \quad \text{and} \quad E_4 (\bp) = 2 t_0 +U_0 - \frac{U_y}{4},
\nn
& \phi_5 (\bp) = \frac{1}{\sqrt 2}(0,0,0,1,-1) \quad \text{and} \quad E_5 (\bp) = 2 t_0 +U_0 - \frac{U_y}{4}.
\ea
For the parameter values, $(U_0, U_x, U_y) = (0.125, 0.10, 0.06)$, the energy eigenvalues are
\bg
E_1 (\bp) \in [1.92, 2.08],
\quad
E_2 (\bp) = E_3 (\bp) = 2.10,
\quad
E_4 (\bp) = E_5 (\bp) = 2.11.
\eg
Thus, $\phi_1 (\bp)$ forms the lowest exciton band.

We now remove the truncation and allow $\bD$ to take any value in $\Z^2$.
For any lattice vector $\bb v \in \Z^2$ other than $\{0,+\hat{x},-\hat{x},+\hat{y},-\hat{y}\}$, we define the state $\phi^{(\bb v)} (\bp)$ through its components, $\phi^{(\bb v)}_\bD (\bp) = \delta_{\bD, \bb v}$.
It follows directly from Eq.~\eqref{seq:H_5x5} that
\bg
\sum_{\bD'} \, \mc H_{\bD, \bD'} \phi^{(\bb v)}_{\bD'} (\bp) = (2 t_0 + U_0) \delta_{\bD, \bb v} = (2 t_0 + U_0) \phi^{(\bb v)}_\bD (\bp).
\eg
Therefore, each $\bb v$ defines an exact eigenstate with energy $E_{\rm cont} = 2 t_0 + U_0 = 2.125$.
The states associated with all larger relative separations span an infinitely degenerate sector.
In this exactly solvable limit, the continuum collapses to the flat energy level $E_{\rm cont}$.
Additional terms coupling states at larger relative separations generically broaden this level into a continuum.

Since $E_{\rm cont} > E_{1,\dots,5} (\bp)$, this sector lies above all five exciton bands retained in the truncated basis.
The truncation therefore captures the low-energy exciton bands, provided that the additional terms introduced beyond the atomic limit are sufficiently small not to close this energy separation.
\\

\subsection{Oblique-lattice realization}
\label{subapp:lattice}
\begin{figure*}[t]
\centering
\includegraphics[width=0.9\textwidth]{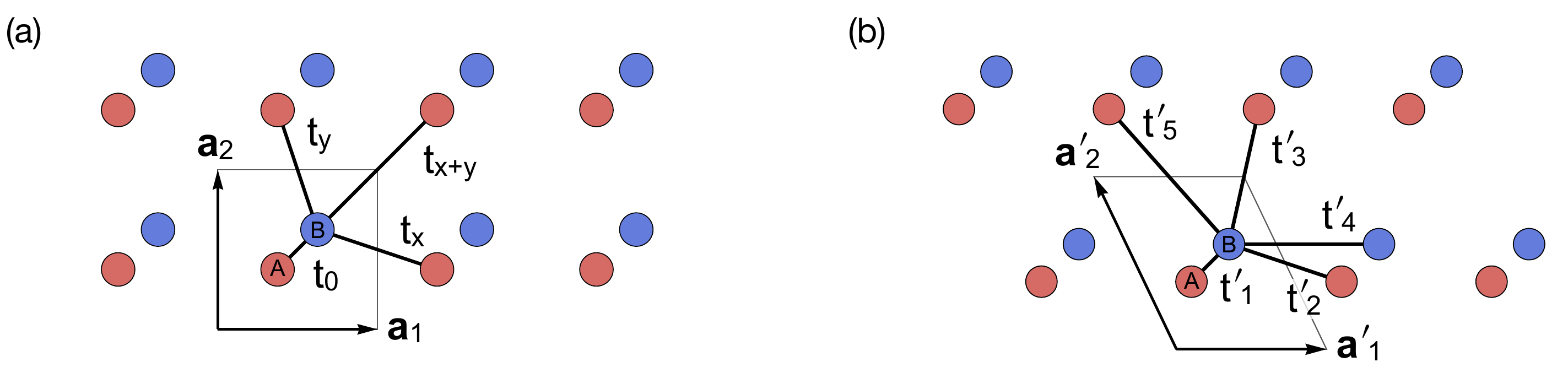}
\caption{
(a) Rectangular lattice used throughout the main text.
The primitive lattice vectors are $\bb a_1 = (1,0)$ and $\bb a_2 = (0,1)$.
Although one may simply choose the sublattice positions $\bx_A = \bx_B = \bb 0$, we instead take $\bx_A= -\bb l/2$ and $\bx_B = \bb l/2$, where $\bb l = (0.25,0.25)$, to facilitate comparison with the oblique-lattice realization in panel (b).
The hopping amplitudes used in the main text are $t_0=1$, $t_x=0.025$, $t_{x+y}=0.02$, and $t_y=0$.
Symmetry-related hoppings are omitted for clarity.
(b) An oblique-lattice realization of the same hopping pattern.
One convenient choice is $\bb a'_1 = (1,0)$ and $\bb a'_2 = (-0.55,1.15)$, and the sublattice positions are $\bx_A = -\bb l/2$ and $\bx_B = \bb l/2$, with the same $\bb l = (0.25,0.25)$.
The hoppings $t'_1,\dots,t'_5$ denote the first- through fifth-nearest-neighbor hoppings, respectively.
The hopping parameters used in the main text correspond to $t_0 = -t'_1$, $t_x = t'_2$, $t_{x+y} = t'_3$, and $t_y = t'_5$, providing a natural interpretation of the model in terms of short-range hoppings on an oblique lattice.}
\label{sfig:lattice}
\end{figure*}

As a concrete realization, we consider the oblique lattice shown in Fig.~\ref{sfig:lattice}(b), where primitive lattice vectors $\bb a'_1 = (1,0)$, $\bb a'_2 = (-0.55, 1.15)$.
The two sublattices within each unit cell are located at $\bx_A = - \bb l /2$ and $\bx_B = \bb l /2$, where $\bb l = (0.25,0.25)$.
The first few hopping distances are $|\bb l| = 0.3536$, $|\bb a'_1 - \bb l| = 0.7906$, $|\bb a'_1 + \bb a'_2 - \bb l| = 0.9220$, $|\bb a'_1|=1$, and $|\bb a'_2 - \bb l| = 1.2042$, corresponding to the hopping amplitudes $t'_1, \dots, t'_5$ in ascending order of distance.
These reproduce the rectangular-lattice model through the correspondence $t_0 \leftrightarrow -t'_1$, $t_x \leftrightarrow t'_2$, $t_{x+y} \leftrightarrow t'_3$, and $t_y \leftrightarrow t'_5$.
The minus sign in the first correspondence is a matter of convention.
The corresponding hopping operators transform accordingly, for example,
\bg
- t_0 \, c^\dg_{\bR, A} \, c_{\bR, B} + h.c. \leftrightarrow t'_1 \, c^\dg_{\bR, A} \, c_{\bR, B} + h.c.,
\nn
t_{x+y} \, c^\dg_{\bR + \hat x + \hat y, A} \, c_{\bR, B} + h.c. \leftrightarrow t'_3 \, c^\dg_{\bR + \bb a_1 + \bb a_2, A} \, c_{\bR, B} + h.c..
\eg
Since the primitive lattice vectors are no longer aligned with $\hat x$ and $\hat y$, it is convenient to introduce momentum coordinates with respect to the reciprocal lattice vectors $\bb g_{1,2}$, $\bp = p_1 \bb g_1 + p_2 \bb g_2$ where $p_{1,2} \in (-1/2,1/2]$.
The high-symmetry momenta are then $\Gamma = \bb 0$, $X = \bb g_1/2$, $Y = \bb g_2/2$, and $M = \bb g_1/2 + \bb g_2/2$.
Replacing $(\hat x,\hat y)$ by $(\bb a_1, \bb a_2)$ and substituting $(p_x, p_y) \to 2 \pi (p_1, p_2)$, the momentum-space Hamiltonian $\mc H (\bp)$ retains exactly the same functional form as in the rectangular-lattice model.

\subsection{Emergent spinless time-reversal symmetry}
\label{subapp:trs}
We begin with the simplest hopping process already introduced in the atomic-limit Hamiltonian,
\bg
t \, \sum_\bR \, c^\dg_{\bR,A} \, c_{\bR,B},
\eg
where, for the moment, $t \in \C$.
Under inversion $\mc I$,
\bg
\mc I \, c^\dg_{\bR,A} \, \mc I^{-1} = c^\dg_{-\bR, B},
\quad
\mc I \, c^\dg_{\bR,B} \, \mc I^{-1} = c^\dg_{-\bR,A},
\eg
so the inversion-transformed term is
\bg
t \, \sum_\bR \, c^\dg_{-\bR,B} \, c_{-\bR,A}
= t \, \sum_\bR \, c^\dg_{\bR,B} \, c_{\bR,A}.
\eg
Therefore, inversion symmetry requires that both terms be present in the Hamiltonian.
Imposing Hermiticity on their sum then gives
\bg
2 {\rm Re} [t] \, \sum_\bR \, \left( c^\dg_{\bR,A} \, c_{\bR,B} + c^\dg_{\bR,B} \, c_{\bR,A} \right).
\eg
Absorbing the overall coefficient into the definition of the hopping amplitude, the Hamiltonian may equivalently be written as
\bg
t \, \sum_\bR \, \left( c^\dg_{\bR,A} \, c_{\bR,B} + c^\dg_{\bR,B} \, c_{\bR,A} \right),
\eg
where $t \in \R$.
Since the hopping amplitude $t$ is real, this term is invariant under the spinless time-reversal operator $\hat T$ defined by
\bg
\hat T \, i \, \hat T^{-1} = -i,
\quad
\hat T \, c^\dg_{\bR,A/B} \, \hat T^{-1} = c^\dg_{\bR,A/B}.
\eg

\tocless{\subsubsection{General hopping terms}}{}
We next consider more general hopping processes.
For inter-sublattice hoppings,
\bg
t \, \sum_\bR \, c^\dg_{\bR + \bb X,A} \, c_{\bR,B},
\eg
where $\bb X$ is a lattice vector and, for the moment, $t \in \C$.
Under inversion, this term is mapped to
\bg
t \, \sum_\bR \, c^\dg_{\bR,B} \, c_{\bR + \bb X,A}.
\eg
Including both terms and imposing Hermiticity gives
\bg
2 {\rm Re} [t] \, \sum_\bR \, \left( c^\dg_{\bR + \bb X,A} \, c_{\bR,B} +c^\dg_{\bR,B} \, c_{\bR + \bb X,A} \right),
\eg
where the hopping amplitude may again be chosen to be real.
Therefore, the microscopic Hamiltonian possesses $T$.

The situation is different for intra-sublattice hoppings of the form $\sum_\bR \, c^\dg_{\bR + \bb X, A} \, c_{\bR, A}$.
For $\bb X = \bb 0$, following the same argument as above, one obtains
\bg
t \, \sum_\bR \, \left( c^\dg_{\bR,A} \, c_{\bR,A} + c^\dg_{\bR,B} \, c_{\bR,B} \right),
\eg
where $t \in \R$, so this term is also invariant under $T$.
For $\bb X \neq \bb 0$, however, after imposing inversion symmetry and Hermiticity, one obtains
\bg
t \, \sum_\bR \, \left( c^\dg_{\bR+\bb X,A} \, c_{\bR,A} + c^\dg_{\bR,B} \, c_{\bR+\bb X,B} \right) + t^* \, \sum_\bR \, \left( c^\dg_{\bR,A} \, c_{\bR+\bb X,A} + c^\dg_{\bR+\bb X,B} \, c_{\bR,B} \right),
\eg
where the hopping amplitude $t$ remains an arbitrary complex number.
Therefore, this hopping process breaks $T$.

Remarkably, after projection onto the exciton basis, this hopping contributes
\bg
{\rm Re} [t] (e^{-i \bp \cdot \bb X} - 1) \delta_{\bD, \bD' + \bb X} + {\rm Re} [t] (e^{i \bp \cdot \bb X} - 1) \delta_{\bD + \bb X, \bD'}
\eg
to the projected Hamiltonian $\mc H_{\bD, \bD'} (\bp)$.
The imaginary part of $t$ drops out completely, so the projected Hamiltonian depends only on ${\rm Re}[t]$.
Equivalently, $\mc H_{\bD, \bD'} (\bp) = \left( \mc H_{\bD, \bD'} (-\bp) \right)^*$ which is precisely the spinless time-reversal condition.
Consequently, no quadratic hopping process can break the emergent time-reversal symmetry $T$.

\tocless{\subsubsection{Density-density interactions}}{}
We next consider density-density interactions.
The most general inter-sublattice density-density interaction takes the form
\bg
U \, \sum_\bR \, n_{\bR+\bb X,A} \, n_{\bR,B},
\eg
where $\bb X$ is a lattice vector.
Likewise, the most general intra-sublattice density-density interaction is
\bg
U' \, \sum_\bR \,\left( n_{\bR+\bb X,A} \, n_{\bR,A} + n_{\bR,B} \, n_{\bR+\bb X,B} \right).
\eg
Since both of the interactions are written entirely in terms of Hermitian number operators, Hermiticity requires the interaction strength $U$ to be real.
Consequently, these interactions are invariant under $T$.

After projection onto the exciton basis, the projected Hamiltonian therefore also preserves the emergent spinless time-reversal symmetry $T$.
Hence, arbitrary density-density interactions preserve $T$.
Combining this result with the previous analysis, we conclude that arbitrary quadratic hopping terms together with arbitrary density-density interactions cannot break the emergent time-reversal symmetry.
For this reason, our model employs interaction terms beyond the density-density type.
\\

\section{A model-building scheme for multiband systems}
\label{app:multiband}
Our real-space construction extends to multiband systems with more than two orbitals per unit cell.
Consider a unit cell containing $N_{\rm orb}$ atomic orbitals with creation operators $a^\dg_{\bR,i}$ $(i=1, \dots, N_{\rm orb})$.
We first define the Wannier states for the conduction and valence bands.
To this end, let $M$ be a $2 \times N_{\rm orb}$ matrix whose rows are orthonormal, such that $M M^\dg = \mathds{1}_2$, with $\mathds{1}_2$ the $2 \times 2$ identity matrix.

The Wannier creation operators are then defined by
\bg
b^\dg_{\bR,l} = \sum_{i=1}^{N_{\rm orb}} \, M_{li} \, a^\dg_{\bR,i},
\eg
where $l=c,v$, and $b^\dg_{\bR,c} \equiv c^\dg_\bR$ and $b^\dg_{\bR,v} \equiv v^\dg_\bR$.
These are also compactly localized states~\cite{read2017compactly,rhim2019classification} since each Wannier state is constructed from atomic orbitals localized within a finite spatial region.
Their canonical anticommutation relations follow directly from those of $a^\dg_{\bR,i}$ together with the orthonormality of the rows of $M$:
\bg
\{ b_{\bR,l}, b^\dg_{\bR', l'} \}
= \sum_{i,j} \, (M^\dg)_{il} M_{l' j} \{ a_{\bR,i}, a^\dg_{\bR',j} \}
= \delta_{\bR, \bR'} (M M^\dg)_{l l'}
= \delta_{\bR,\bR'} \delta_{l,l'}.
\eg

A parent atomic Hamiltonian realizing the chosen conduction and valence Wannier orbitals can always be constructed.
For example, let
\bg
U= \bpm
M \\ W
\epm,
\eg
be an $N_{\rm orb} \times N_{\rm orb}$ unitary matrix, where the rows of $W$ span the orthogonal complement of the row space of $M$.
Then, $M$ and $W$ satisfy
\bg
W W^\dg = \mathds{1}_{N_{\rm orb}-2},
\quad
W M^\dg = 0,
\quad
M^\dg M + W^\dg W = \mathds{1}_{N_{\rm orb}}.
\eg
Using $W$, we define creation operators for the remaining atomic-limit bands,
\bg
d^\dg_{\bR,\mu} = \sum_i \, W_{\mu i} \, a^\dg_{\bR,i}
\quad
(\mu = 1, \dots, N_{\rm orb}-2).
\eg
The inverse transformation is
\bg
a^\dg_{\bR,i} = \sum_{l=c,v} \, M_{li}^* \, b^\dg_{\bR,l} + \sum_\mu \, W_{\mu i}^* \, d^\dg_{\bR,\mu}.
\eg

Choosing the energies $\ep_\mu$ such that all inactive bands are well separated from the active conduction and valence bands compared with the characteristic microscopic hopping and interaction scales, one may construct the parent Hamiltonian
\bg
\hat H_0 = \sum_\bR \, \Big[ \ep_c \, b^\dg_{\bR,c} \, b_{\bR,c} + \ep_v \, b^\dg_{\bR,v} \, b_{\bR,v} + \sum_\mu \, \ep_\mu \, d^\dg_{\bR,\mu} \, d_{\bR,\mu} \Big].
\eg
Expressing $\hat H_0$ in terms of $a^{(\dg)}_{\bR,i}$ yields the corresponding microscopic tight-binding Hamiltonian.
Similar constructions have been employed in various contexts, including flat-band systems~\cite{graf2021designing,hwang2021general}.
To describe excitons predominantly formed from a selected conduction-valence pair, we restrict the Hilbert space to the corresponding active conduction-valence subspace.
The subsequent Hamiltonian is formulated directly within this active subspace, where additional hopping and interaction terms may be introduced as effective Hamiltonian parameters to capture the exciton physics of interest.
We then define the reference state $\gsk$ as the filled-valence state within this subspace.
Here, $\gsk$ serves as the reference state for the second-quantized representation of electron-hole excitations.
We then construct the one-exciton Hilbert space within the active conduction-valence subspace following the procedure developed in Sec.~\ref{app:real_space}.
The subsequent exciton construction proceeds identically within this active conduction-valence subspace.

\section{Connection to the momentum-space formalism}
\label{app:momentum}

\subsection{Momentum-space formulation}
To establish the connection with the conventional momentum-space formulation of excitons, we introduce the corresponding momentum-space basis and derive its relation to the real-space basis developed in Sec.~\ref{app:real_space}.
Consider conduction- and valence-band creation operators $c^\dg_\bk$ and $v^\dg_\bk$.
An exciton basis with hole momentum $\bk$ and total momentum $\bp$ is written as
\bg
\ket{\bk, \bp} = c^\dg_{\bk + \bp} \, v_\bk \gsk.
\eg
Although the conventional momentum-space formalism typically takes the reference ground state as $\prod_\bk \, v^\dg_\bk \ket{0}$, we instead retain the real-space definition $\gsk = \prod_\bR \, v^\dg_\bR \ket{0}$ introduced in Sec.~\ref{app:real_space}.
These two states differ only by an overall $U(1)$ phase factor and are therefore physically equivalent.
Using the same reference state throughout simplifies the mapping between the momentum- and real-space formalisms.

The basis $\ket{\bk, \bp}$ is related to the real-space bases $\ket{\bD, \bR}$ and $\ket{\bD, \bp}$.
Using Eq.~\eqref{seq:wannier}, we obtain
\bg
\ket{\bk, \bp} = \frac{1}{N} \sum_{\bR_1, \bR_2} \, e^{i (\bk+\bp) \cdot \bR_1 - i \bk \cdot \bR_2} \, c^\dg_{\bR_1} \, v_{\bR_2} \gsk
= \frac{1}{N} \sum_{\bR, \bD} \, e^{i \bp \cdot \bR + i \bk \cdot \bD} \, c^\dg_\bR \, v_{\bR-\bD} \gsk.
\eg
Using the definitions of $\ket{\bD, \bR}$ [Eq.~\eqref{seq:basis_real}] and $\ket{\bD, \bp}$ [Eq.~\eqref{seq:basis_partial1}], this becomes
\ba
\ket{\bk, \bp} =& \frac{1}{N} \sum_{\bR, \bD} \, e^{i \bp \cdot \bR + i \bk \cdot \bD} \, \ket{\bD, \bR}
= \frac{1}{\sqrt N} \sum_\bD \, e^{i \bk \cdot \bD} \, \ket{\bD, \bp},
\nn
\ket{\bD, \bp} =&\frac{1}{\sqrt N} \sum_\bk \, e^{-i \bk \cdot \bD} \, \ket{\bk, \bp}.
\label{seq:basis_relation}
\ea

Following the real-space formalism developed in Sec.~\ref{app:real_space}, the projected Hamiltonian corresponding to a Hamiltonian $\hat H$ is defined by
\bg
\mc H_{\bk, \bk'} (\bp) = \bra{\bk, \bp} \hat H \ket{\bk', \bp} - \delta_{\bk, \bk'} \gsb \hat H \gsk.
\label{seq:projH_k}
\eg
Comparing Eq.~\eqref{seq:projH_k} with Eq.~\eqref{seq:projH_p}, we obtain
\bg
\mc H_{\bk, \bk'} (\bp) = \frac{1}{N} \sum_{\bD, \bD'} \, e^{-i \bk \cdot \bD + i \bk' \cdot \bD'} \, \mc H_{\bD, \bD'} (\bp).
\eg
The projected Hamiltonians $\mc H_{\bD,\bD'} (\bp)$ and $\mc H_{\bk,\bk'} (\bp)$ therefore describe the same exciton problem in different bases.
Using Eqs.~\eqref{seq:exc_state} and \eqref{seq:projHop}, the exciton envelope wave functions, eigenstates, and projected Hamiltonian operator can be rewritten as
\ba
\phi_{n, \bD} (\bp) &= \frac{1}{\sqrt N} \sum_\bk \, e^{i \bk \cdot \bD} \phi_{n,\bk} (\bp),
\quad
\phi_{n, \bk} (\bp) = \frac{1}{\sqrt N} \sum_\bD \, e^{-i \bk \cdot \bD} \phi_{n,\bD} (\bp),
\nn
\ket{\Phi_n (\bp)} &= \sum_\bD \, \phi_{n, \bD} (\bp) \ket{\bD, \bp}
= \sum_\bk \, \phi_{n,\bk} (\bp) \ket{\bk, \bp},
\nn
\hat {\mc H} (\bp) &= \sum_{\bD, \bD'} \, \ket{\bD, \bp} \mc H_{\bD, \bD'} (\bp) \bra{\bD', \bp}
= \sum_{\bk, \bk'} \, \ket{\bk, \bp} \mc H_{\bk, \bk'} (\bp) \bra{\bk', \bp}.
\label{seq:fourier}
\ea

Finally, we relate the inversion symmetry representations in the two bases.
From Eq.~\eqref{seq:inv_op3}, we have
\bg
\hat U_{\mc I} (\bp)
= \sum_\bD \, \ket{ -\bD + 2\bx_v - 2\bx_c, -\bp} \, e^{-2 i \bp \cdot \bx_c} \lambda_{\rm GS} \lambda_c (\bb 0) \lambda_v (\bb 0)^{-1} \, \bra{\bD, \bp}
\eg
Using Eq.~\eqref{seq:basis_relation}, this becomes
\bg
\hat U_{\mc I} (\bp)
= \sum_\bk \, \ket{-\bk, -\bp} \, e^{-2 i (\bk+\bp) \cdot \bx_c + 2i \bk \cdot \bx_v} \lambda_{\rm GS} \lambda_c (\bb 0) \lambda_v (\bb 0)^{-1} \, \bra{\bk, \bp}.
\eg
Using the definition of the inversion sewing matrix in Eq.~\eqref{seq:inv_sewing}, the above equation can be rewritten as
\bg
\hat U_{\mc I} (\bp)
= \sum_\bk \, \ket{-\bk, -\bp} \lambda_{\rm GS} \lambda_c (\bk + \bp) \lambda_v (\bk)^{-1} \bra{\bk, \bp}
\eg
Accordingly, the exciton symmetry relation [Eq.~\eqref{seq:exc_inversion1}] becomes (omitting the exciton index $n$)
\bg
\lambda_c (\bk+\bp) \lambda_v (\bk)^{-1} \phi_\bk (\bp) = \lambda_{\rm exc} (\bp) \phi_{-\bk} (-\bp).
\label{seq:exc_inversion_k}
\eg
This is the symmetry constraint obtained by specializing the general symmetry formalism of Refs.~\cite{nalabothula2026symmetries,hwang2026stable} to inversion symmetry.
For convenience, we reproduce Eq.~\eqref{seq:exc_inversion2} here:
\bg
e^{-2 i \bp \cdot \bx_c} \, \lambda_c (\bb 0) \lambda_v (\bb 0)^{-1} \, \phi_{\bD} (\bp) = \lambda_{\rm exc} (\bp) \, \phi_{-\bD + 2\bx_v - 2\bx_c} (-\bp).
\label{seq:exc_inversion_r}
\eg

\subsection{Stable zeros in momentum space}
Let us restrict both $\bk$ and $\bp$ to inversion-invariant momenta $\cm \bk$ and $\cm \bp$, satisfying $\cm \bk = - \cm \bk$ and $\cm \bp = - \cm \bp$ up to reciprocal lattice vectors.
Thus, $\cm \bk, \cm \bp \in \{\Gamma, X, Y, M\}$, where $\Gamma=0$, $X= \frac{1}{2} \bb g_1$, $Y = \frac{1}{2} \bb g_2$, and $M = \frac{1}{2} \bb g_1 + \frac{1}{2} \bb g_2$, with primitive reciprocal lattice vectors $\bb g_{1,2}$.
The symmetry constraints in Eqs.~\eqref{seq:exc_inversion_k} and \eqref{seq:exc_inversion_r} reduce to
\bg
\lambda_c (\cm \bk + \cm \bp) \lambda_v (\cm \bk)^{-1} \phi_{\cm \bk} (\cm \bp) = \lambda_{\rm exc} (\cm \bp) \phi_{\cm \bk} (\cm \bp),
\nn
e^{- 2 i \cm \bp \cdot \bx_c} \lambda_c (\bb 0) \lambda_v (\bb 0)^{-1} \phi_{\bD} (\cm \bp) = \lambda_{\rm exc} (\cm \bp) \phi_{-\bD + 2\bx_v - 2\bx_c} (\cm \bp).
\eg
When $\lambda_c (\cm \bk + \cm \bp) \lambda_v (\cm \bk)^{-1}$ and $\lambda_{\rm exc} (\cm \bp)$ mismatch, the stable zero $\phi_{\cm \bk} (\cm \bp) = 0$ is enforced~\cite{hwang2026stable}.
In the presence of $n$-fold rotation symmetry ($n=2,3,4,6$), the exciton Chern number modulo $n$ is determined by the rotation eigenvalues $\lambda_{\rm exc} (\cm \bp)$~\cite{fang2012bulk,hwang2026stable}.
For inversion symmetry, this reduces to the twofold rotation case.

Since both of $\phi_{\cm \bk} (\cm \bp)$ and $\phi_\bD (\cm \bp)$ represent the same exciton eigenstate in different spaces, they share the same rotation eigenvalue $\lambda_{\rm exc} (\cm \bp)$.
Consequently, the exciton Chern number modulo $n$ may equally be inferred from the real-space envelope wave function.
More importantly, the real-space formalism provides a direct connection between the symmetry of $\phi_\bD (\cm \bp)$ and momentum-space stable zeros, as we explain below.

Ref.~\cite{hwang2026stable} also established a correspondence between the stable-zero configuration at high-symmetry momenta and the topology of the exciton and the underlying bands.
We now establish the corresponding real-space characterization in terms of the symmetry of $\phi_\bD (\cm \bp)$.
Let us denote the number of stable zeros at $(\bk, \bp) = (\cm \bk, \cm \bp)$ by $N_{(\cm \bk, \cm \bp)} \in \{0, 1\}$.
For inversion (or equivalently twofold rotation) symmetry, the Chern numbers modulo $2$ of the exciton and the underlying bands are given by
\ba
C_{\rm exc} &= N_{(\Gamma, \Gamma)} + N_{(\Gamma, X)} + N_{(Y,Y)} + N_{(Y,M)} \pmod 2,
\nn
C_c &= N_{(\Gamma, Y)} + N_{(Y,Y)} + N_{(\Gamma, M)} + N_{(Y,M)} \pmod 2,
\nn
C_v &= N_{(\Gamma,X)} + N_{(X,X)} + N_{(Y,M)} + N_{(M,M)} \pmod 2.
\label{seq:counting}
\ea
The above relations show that the topology can be inferred entirely from the stable-zero configuration.
In the following, we establish the corresponding real-space criterion by relating the parity of $\phi_\bD (\cm \bp)$ to the stable zeros.

In our model, $\bx_v=\bx_c=\bb 0$ and $(\lambda_v (\bb 0), \lambda_c (\bb 0))=(+1,-1)$.
Equation~\eqref{seq:exc_inversion_r} therefore reduces to
\bg
\phi_\bD (\cm \bp) = - \lambda_{\rm exc} (\cm \bp) \phi_{-\bD} (\cm \bp).
\label{seq:constraint_model}
\eg
The inversion eigenvalues $\lambda_{\rm exc} (\cm \bp)$ of the lowest exciton at $(\Gamma,X,Y,M)$ are $(-1,-1,-1,+1)$ implying the exciton Chern number $C_{\rm exc} = 1$ modulo 2.
(As discussed in the main text and Sec.~\ref{app:model}, we absorbed the overall minus sign into the definition of the exciton inversion eigenvalue and therefore denoted the inversion eigenvalues as $(+1,+1,+1,-1)$ there.)

We first consider $\cm \bp = M$.
Since $\lambda_{\rm exc} (\bp) = +1$, Eq.~\eqref{seq:constraint_model} gives $\phi_\bD (\cm \bp) = - \phi_{-\bD} (\cm \bp)$, showing that $\phi_\bD (\cm \bp)$ is odd under inversion.
Since $\bD = \bb 0 \in \Z^2$ is the unique inversion-invariant lattice point, odd parity immediately implies $\phi_{\bD = \bb 0} (\cm \bp) = 0$.
Using Eq.~\eqref{seq:fourier},
\bg
\sqrt N \phi_{\cm \bk} (\cm \bp)
= \phi_{\bD = \bb 0} (\cm \bp) + \sum_{\bD \ne \bb 0} \, e^{- i \cm \bk \cdot \bD} \phi_\bD (\cm \bp).
\eg
The first term vanishes by odd parity.
The remaining terms can be grouped into inversion-related pairs $(\bD, -\bD)$, each contributing $\sin (\cm \bk \cdot \bD) \phi_\bD (\cm \bp)$.
Since $\cm \bk \cdot \bD \in \pi \Z$ for every inversion-invariant momentum $\cm \bk$, these contributions also vanish.
Hence, $\phi_{\cm \bk} (\cm \bp) = 0$ for all $\bk \in \{\Gamma, X, Y, M\}$.

For $\cm \bp = \Gamma, X, Y$, we have $\lambda_{\rm exc} (\bp) = -1$, so that $\phi_\bD (\cm \bp) = \phi_{-\bD} (\cm \bp)$.
The Fourier amplitude $\phi_{\cm \bk} (\cm \bp)$ is therefore not constrained to vanish.
Consequently, $N_{(\cm \bk, M)}=1$ for all $\cm \bk \in \{\Gamma, X, Y, M\}$, while $N_{(\cm \bk, \cm \bp)}=0$ for all other cases.
Then, Eq.~\eqref{seq:counting} immediately yields
\bg
C_{\rm exc} = 1 \pmod 2,
\quad
C_c = C_v = 0 \pmod 2,
\eg
in agreement with the momentum-space characterization.

Finally, we comment on the case where $\bx_{c,v}$ are nontrivial, for example $\bx_c = (0,0)$ and $\bx_v = (1/2,0)$.
In this case, Eq.~\eqref{seq:exc_inversion_r} becomes $\phi_\bD (\cm \bp) = -\phi_{-\bD + 2\bx_v - 2\bx_c} (\cm \bp)$.
One might expect that $\phi_\bD (\cm \bp)$ must vanish whenever $-\bD + 2\bx_v - 2\bx_c = \bD$.
However, such a lattice vector $\bD$ need not exist.
For example, when $\bx_c = (0,0)$ and $\bx_v = (1/2,0)$, the fixed-point condition gives $\bD = (1/2,0)$ which is not an element of $\Z^2$.
Taking this subtlety into account, the resulting real-space criterion again reproduces the momentum-space criterion of Ref.~\cite{hwang2026stable}.

\section{Exciton Wilson loop}
\label{app:wilson}
In Ref.~\cite{davenport2026exciton}, Wilson loops for excitons were established.
Two exciton Berry connections can be defined, corresponding to the electron and hole sectors, respectively, by projecting the position operator onto the conduction- and valence-band subspaces.
Exponentiating the corresponding Berry connections yields two exciton Wilson loops, both of which give the same Chern number.
We therefore consider only the Wilson loop associated with the conduction-band sector.

The Wilson loop is constructed as the ordered product of Wilson lines, each defined between two nearby momenta $\bp$ and $\bp'$:
\bg
\mc W^{(c)}_{\bp' \leftarrow \bp}
= \sum_\bk \, \brk{u_{\bk + \bp',c}}{u_{\bk + \bp,c}} \left( \phi_\bk (\bp') \right)^* \phi_\bk (\bp),
\label{seq:wilson_def}
\eg
where $u_{\bk,c}$ and $u_{\bk,v}$ denote the Bloch or tight-binding eigenstates of the conduction and valence bands, respectively.

We now express the Wilson line in Eq.~\eqref{seq:wilson_def} in terms of the real-space envelope wave function $\phi_\bD (\bp)$.
To this end, we express the Wannier creation operators $c^\dg_{\bR,l}$ ($l=c,v)$ in terms of the atomic orbital creation operators as
\bg
c^\dg_{\bR, l} = \sum_{\bR', i} \, S_{l i} (\bR') \, c^\dg_{\bR+\bR', i}.
\eg
Here, $c^\dg_{\bR,i}$ creates an electron in the atomic orbital $i$ of the unit cell at $\bR$, and $S_{li}(\bR)$ specifies the compact Wannier state in the atomic-orbital basis~\cite{read2017compactly,hwang2021flat,schindler2021noncompact}.
The orthonormality of the Wannier basis implies $\sum_{\bR'', i} \, \left( S_{l i}(\bR'' - \bR) \right)^* S_{l' i} (\bR'' - \bR') = \delta_{l,l'} \delta_{\bR,\bR'}$.
Furthermore, letting $\bx_i$ denote the position of orbital $i$ within the unit cell, one finds
\bg
(u_{\bk, l})_i = \sum_\bR \, S_{l i} (\bR) e^{-i \bk \cdot (\bR + \bx_i)}.
\label{seq:u_wannier}
\eg
Substituting Eq.~\eqref{seq:u_wannier} into Eq.~\eqref{seq:wilson_def}, we obtain
\ba
\mc W^{(c)}_{\bp' \leftarrow \bp}
=& \sum_{\bR, \bD, \bD', i} \, \left( \phi_\bD (\bp') \right)^* \phi_{\bD'} (\bp) \left( S_{c,i} (\bR-\bD) \right)^* S_{c,i} (\bR-\bD') e^{-i \bp \cdot (\bD-\bD')} e^{i (\bp'-\bp) \cdot (\bR - \bD + \bx_i)}.
\label{seq:model_wl}
\ea

We now apply this expression to the model studied in the main text.
The conduction-band Wannier operator is $c^\dg_\bR = \frac{1}{\sqrt 2} (c^\dg_{\bR,A} - c^\dg_{\bR,B})$ [Eq.~\eqref{seq:model_basis}], and hence $S_{c,A} (\bR) = \frac{1}{\sqrt 2} \delta_{\bR,\bb 0}$ and $S_{c,B} (\bR) = - \frac{1}{\sqrt 2} \delta_{\bR,\bb 0}$.
In Fig.~\ref{sfig:lattice}, the positions $\bx_A$ and $\bx_B$ are spatially separated only to illustrate the hopping structure.
For evaluating the Wilson loop, we choose the coincident embedding $\bx_A=\bx_B=\bb 0$.
This choice does not affect the energy spectrum or the inversion eigenvalues of the model.
Moreover, the orbital positions may be continuously deformed to this coincident embedding without closing the exciton gap, and therefore without changing the Chern number or the winding of the Wilson loop.
With this choice, Eq.~\eqref{seq:model_wl} reduces to
\ba
\mc W^{(c)}_{\bp' \leftarrow \bp}
=& \sum_{\bD, i} \, \left( \phi_\bD (\bp') \right)^* \phi_\bD (\bp) \left| S_{c,i} (\bb 0) \right|^2
= \sum_\bD \, \left( \phi_\bD (\bp') \right)^* \phi_\bD (\bp).
\ea
For each fixed $p_y$, the Wilson loop along the $p_x$ direction is evaluated as the product
\bg
\mc W (p_y) =
\prod_{p_x=0}^{2\pi-\delta p} \, \sum_\bD \, \left( \phi_\bD (p_x+\delta p, p_y) \right)^* \phi_\bD (p_x,p_y)
\eg
where $\delta p$ is the momentum spacing along the discretized $p_x$ direction.
The Wilson-loop phases $\arg \mc W (p_y) \in [-\pi,\pi)$ are then plotted as functions of $p_y \in [-\pi,\pi)$.
This is the procedure used to obtain the Wilson-loop spectrum shown in the main text.

\end{document}